\documentclass[11pt]{article}
\usepackage{graphicx}
\usepackage{amssymb}

\newtheorem{lemma}{Lemma}
\newtheorem{proposition}{Proposition}

\def\proof{\noindent  {\underline {Proof}}. }
\def\square{ {\hfill \vrule height6pt width6pt depth1pt} \bigskip \medskip }

\def\proof{\noindent  {\underline {Proof}}. }
\def\square{ {\hfill \vrule height6pt width6pt depth1pt} \newline  }

\def\today{\number\day\space\ifcase\month\or Janvier \or F\'evrier \or  Mars 
   \or Avril \or Mai \or Juin \or Juillet \or Ao\^ut \or Septembre \or Octobre 
   \or Novembre \or D\'ecembre \fi\number \year}

\begin{document}

\centerline{\Large Continuum limit of the Volterra model, separation of variables }
\centerline{\Large and non standard realizations of the Virasoro Poisson bracket.}

\vskip 1cm
\centerline{O. Babelon
\footnote{Member of CNRS.}}
\vskip .5cm
\centerline{Laboratoire de Physique Th\'eorique et Hautes Energies\footnote{Tour 24-25, 5\` eme \'etage, Boite 126, 
4 Place Jussieu,  75252 Paris Cedex 05.
 }, (LPTHE)}
\centerline{Unit\'e Mixte de Recherche (UMR 7589)}
\centerline{Universit\'e Pierre et Marie Curie-Paris6; CNRS; Universit\'e Denis Diderot-Paris7;}

\vskip 5cm

\begin{abstract}
The classical Volterra model, equipped with the Faddeev-Takhtadjan
Poisson bracket provides a lattice version of the Virasoro algebra.
The Volterra model being  integrable, we can express the dynamical variables
in terms of the so called separated variables. 
Taking the continuum limit of these formulae,
we obtain the Virasoro generators written
as determinants of  infinite matrices, the elements of which 
are constructed with a set of points lying on an infinite genus Riemann surface.
The coordinates of these points are  separated variables for an 
infinite set of Poisson commuting quantities including $L_0$.
The scaling limit of the eigenvector can also be calculated explicitly, 
so that the associated  Schroedinger equation is in fact exactly solvable.
\end{abstract}

\vfill

\eject

\section{Introduction.}

The relation between integrable systems and conformal field theory has long been recognized
\cite{Gervais85, BaZaLu}. Although the emphasis has been put rightfully  on Baxter $Q$ operator
 and therefore  on Sklyanin's separated variables \cite{Skly85}, to the best of our knowledge
 there is no explicit expressions of the Virasoro generators in terms of these variables.
 
 We make here a  first step in this direction 
by considering the classical version of this problem.  Our strategy will be to start with the
 Volterra model on the lattice \cite{KvM, DuKrNo90} equipped with the Faddeev-Takhtadjan \cite{FaTa86, Volkov86}
Poisson bracket. Since the Volterra model is integrable, we 
can rewrite everything  in terms of separated variables. Now, the Faddeev-Takhtadjan
bracket goes directly to the Virasoro Poisson bracket in the continuum limit, 
and therefore by taking this limit in the separated variables formulae we will obtain the 
Virasoro generators expressed  in terms of separated variables.

This leads to the following rather new type of formula for the Virasoro generators:
$$
u(x) = \sum L_n e^{2in\pi x}   = p_0^2 + 2 \partial_x^2 \log \det \Theta(x) + (L_0 -p_0^2)\delta(x)
$$
Here $p_0$ is the zero mode and Poisson commutes with everything, the term $(L_0 -p_0^2) \delta(x)$ will be explained later and the formula for $L_0$ is given in eq.(\ref{defL0}). 
The infinite matrix  $\Theta(x)$ reads ($k,m \in \{ 1,\cdots , \infty \}$):
 \begin{equation}
\Theta_{km}(x) = {W_k(x) \partial_xE_{m}(x) - \partial_xW_k(x) E_{m}(x) \over Z_k^2 - m^2 \pi^2 },
\quad 0\leq x \leq 1
\label{thetax1}
\end{equation}
with 
\begin{equation}
W_k(x) =   {\sin Z_k x \over Z_k } 
+ \mu_k {\sin Z_k(1- x) \over Z_k },\quad
E_m(x) = 2m\pi \sin m\pi x
\label{WE}
\end{equation}
The above formula for $u(x)$ is valid on the interval $0 \leq x \leq 1$, and should be extended
outside this interval by periodicity (in particular the $\delta(x)$ term in a Dirac comb).

The result of this paper is that if the variables $Z_k, \mu_k$, have Poisson bracket
\footnote{
Notice that if we redefine $\Lambda_k = \sqrt{ Z_k^2 - p_0^2 }$, the Poisson bracket 
becomes a standard quadratic bracket  $\{ \Lambda_k , \mu_k \} = 2 \Lambda_k \mu_k$.
However $p_0$ will then enter the formula for $\Theta(x)$ and, in this work, we prefer to keep that formula  
simple at the expense of  a slightly more complicated Poisson bracket.}
\begin{equation}
\{ Z_k, Z_{k'} \} = 0 , \quad 
\{ Z_k, \mu_{k'} \} = 2 ( Z_k - p_0^2 Z_k^{-1} ) \mu_k  \delta_{kk'} , \quad
\{ \mu_k, \mu_{k'} \} = 0
\label{poisson}
\end{equation}
then $u(x)$ does satisfies the Virasoro Poisson bracket: 
\begin{equation}
\{ u(x), u(y) \} = 4 (u(x)+u(y)) \; \delta'(x-y) + 2 \delta'''(x-y)
\label{virpois}
\end{equation}
Morever, the variables 
$Z_k, \mu_k$ are separated variables for an infinite set of higher commuting quantities, including $L_0$.

Since the separated variables are also the ones which solve the classical 
inverse problem, the Schroedinger equation with the potential $u(x)$
$$
(-\partial_x^2 - u(x))\psi(x, \Lambda)= \Lambda^2 \psi(x,\Lambda))
$$
is exactly solvable, meaning that we have explicit formulae for both the potential $u(x)$ and a basis 
of solutions $\psi(x,\Lambda)$. Constructing the linear combination which is quasi periodic 
(the so called Bloch waves) introduces an infinite genus Riemann surface. The coefficients
in the expression of this curve define a complete set of Poisson commuting Hamiltonians including 
$L_0$. The separated variables are points on this curve.

The paper is organized as follows. In the first three sections we recall some known facts about the 
Volterra model on the lattice. In particular we recall the formulae expressing the dynamical degrees of
freedom in terms of the separated variables. 

In section 5 we compute the continuum (scaling) limit
of the spectral curve. The result is eq.(\ref{specurveH0Hm}). We then show that the Hamiltonians $H_m$
in this formula are in involution.  Moreover we show that the scaling limit of the dynamical divisor 
still belongs to that  curve, and hence define separated variables for these Hamiltonians.

In sections 6, 7 and 8, we compute the scaling limit of the eigenvector of the Lax matrix at each point 
of the spectral curve.  The result is rather simple and is given in eq.(\ref{baker1}). We then show  
that the obtained  expression does satisfy a second order Schroedinger equation  and we compute its potential $u(x)$.   Finally, we construct the two quasi periodic solutions of that equation, the Bloch waves, and 
recover in this way exactly the same spectral curve as the one obtained in section 5.

In section 9,  we give conditions under which the determinants  of the infinite  matrices that appeared
in the previous sections exist.  We then perform a few checks in a certain perturbative scheme. In section 10 we prepare the ground for the  serious calculations coming next.

In sections 11 and 12 we prove that the potential $u(x)$ does satisfy the Virasoro Poisson 
bracket. An essential use is made  of certain quartic relations, proven very much like the
 Hirota--Sato bilinear identities. These identities should be considered as generalizations for $\tau$-functions of the  quartic relations on Riemann's Theta functions.

\section{The Volterra model.}
In this and the following two sections we recall some well known facts about the 
Volterra model.
The Volterra model, as an integrable system, was introduced in \cite{KvM}. It is a restricted
version of the Toda lattice.
We consider a periodic  lattice with $N+1$ sites, and on each lattice site we attach a dynamical
variable $a_i$ on which we impose the Faddeev-Takhtadjan Poisson bracket \cite{FaTa86}:
\begin{equation}
\{a_i,a_j\} =a_ia_j \Big[(4-a_i-a_j)(\delta_{i,j+1}-\delta_{j,i+1})
-a_{j+1}\delta_{i,j+2} + a_{i+1}\delta_{j,i+2} \Big]
\label{fata}
\end{equation}
This bracket\footnote{In terms of Toda Hamiltonian structures, it is  a linear combination  of restrictions of the second and fourth Poisson brackets.} is interesting because taking the continuum limit as
$$
a_i \simeq 1 + \Delta^2 u(x), \quad  \Delta ={1\over (N+1)}
$$
it becomes the Virasoro Poisson bracket eq.(\ref{virpois}).
For precisely this reason, and in this perspective, the lattice model has been extensively 
studied both at the classical level \cite{FaTa86, Volkov86} and at the quantum level \cite{Ba90,Volk1, FaVo, FaKaVo, FaVo2}. The present paper is one more contribution
 this series of works.

The Lax matrix for the Volterra model is defined by:
\begin{eqnarray}
L(\mu) &=& \pmatrix{ 0 & \sqrt{a_1} & 0 & & \cdots& &
{\mu^{-1} }\sqrt{a_{N+1}} \cr\sqrt{a_1} & 0 &
\sqrt{a_2} & & \cdots & & 0 \cr
\vdots & & \ddots & & && \vdots \cr
0 & & \sqrt{a_{i-1}} & 0 & \sqrt{a_i} && 0 \cr
\vdots & & & & \ddots & & \vdots \cr
0 & &\cdots & & \sqrt{a_{N-1}} & 0 & \sqrt{a_N} \cr
{\mu }\sqrt{a_{N+1}} & &\cdots & & 0 & 
\sqrt{a_N} & 0 } \label{ClosedL}
\end{eqnarray}
It is well known that ${\rm Tr} L^n(\mu)$ are in involution with respect to the 
Poisson bracket eq.(\ref{fata}). Hence we have an integrable
system on the lattice whose continuum limit is directly related to 
conformal field theory.

The spectral curve  $\Gamma $  is  defined as usual:
\begin{eqnarray}
\Gamma  \quad :\quad \det (L(\mu) - \lambda)=0 \label{spect} 
\end{eqnarray}
Expanding the determinant we see that it is of the
form:
\begin{eqnarray}
 \mu + \mu^{-1} -  t(\lambda) =0
\label{gamma}
\end{eqnarray}
where  $t(\lambda)$ is  polynomial of degree $N+1$.
\begin{equation}
t(\lambda) ={\cal A}^{-1} \lambda^{N+1} - {\cal A}^{-1}\left(\sum_i a_i\right) \lambda^{N-1} +\cdots 
\label{toflambda}
\end{equation}
where
$$
{\cal A} = \sqrt{a_1 a_2 \cdots a_{N+1}}
$$
Assuming $N=2n$ {\em even},  $t(\lambda)$ is  an {\em odd} polynomial,
$t(-\lambda) = - t(\lambda)$, and has exactly $n + 1$ independent coefficients.
However, in that case,  there is one Casimir function  $K=t(2)$:
$$
\{ t(2), a_i\} = 0, \quad \forall i
$$
The dimension of phase space is $N= 2n$ and we have exactly  $n$ commuting 
quantities. The genus of the curve $\Gamma $ is $g = N$.

At each point $(\lambda,\mu)$ of the spectral curve, we can attach an eigenvector $\Psi(\lambda,\mu) = (\psi_i(x))$, $i=1, \cdots , N+1$,  corresponding to the  eigenvalue $\lambda$ of $L(\mu)$. Explicitly, the equation
$(L(\mu) - \lambda )\Psi = 0$ reads
\begin{eqnarray}
\sqrt{ a_1} \psi_2 + \mu^{-1} 
\sqrt{a_{N+1}} \psi_{N+1} &=& \lambda
\psi_1 \nonumber \\ 
\sqrt{a_{i-1}}\psi_{i-1} 
+ \sqrt{a_i} \psi_{i+1}
&=& \lambda \psi_i \label{recur} \\
\mu \sqrt{a_{N+1}} \psi_1 + \sqrt{a_N} 
\psi_N  &=& \lambda \psi_{N+1} \nonumber
\end{eqnarray}
We extend the definition of the coefficients $a_i$ by periodicity
$a_{i+N+1} = a_i$, and introduce a second order difference
operator $\cal D$:
\begin{eqnarray}
\Big({\cal D} \Psi\Big)_i &\equiv& \sqrt{a_{i-1}}
\psi_{i-1}  + \sqrt{a_i} \psi_{i+1} \nonumber
\end{eqnarray}
This operator is a discrete version of a Schroedinger
operator with periodic potential. Eqs.(\ref{recur}) are then equivalent to:
\begin{eqnarray}
\Big({\cal D} \Psi\Big)_i = \lambda \psi_i, \quad {\rm with}
\quad \psi_{i+N+1} = 
\mu \psi_i \label{schroe} 
\end{eqnarray}
Therefore, the eigenvector $\Psi$ is a Bloch wave for the difference operator $\cal D$ with
 a Bloch momentum $\mu$.  

In the continuum limit, eq.(\ref{schroe}) becomes the Schroedinger equation.
$$
(-\partial_x^2 - u(x)) \psi(x)= \Lambda^2 \psi(x), \quad \lambda \simeq 2 -\Delta^2 \Lambda^2
$$

\section{The free case.}

Since in the continuum limit $a_i \to 1$, it is useful  to first recall some formulae in the trivial case  $a_i = 1$. They will be generalized to the full case in the next section. To introduce the zero mode from the start, we  consider the slightly more general case  $a_i = a$:
\begin{eqnarray*}
\sqrt{a}(\psi_2 + \mu^{-1} \psi_{N+1}) &=& \lambda \psi_1 \\
\sqrt{a}(\psi_{i-1} + \psi_{i+1}) &=& \lambda \psi_i \\
\sqrt{a}(\mu \psi_1 + \psi_N) &=& \lambda \psi_{N+1}
\end{eqnarray*}
The solution of the bulk equations is $\psi_i = \alpha x_+^i + \beta x_-^i$
where $x_\pm$ are solutions of the equation
$$
x^2 - z x + 1 = 0, \quad x_\pm (\lambda) = {1\over 2} (z \pm \sqrt{z^2 -4}), \quad
z={\lambda\over \sqrt{a}}
$$
Imposing the two boundary equations, we get
\begin{eqnarray*}
( x_+^{N+1} -\mu)\; \alpha + ( x_-^{N+1} -\mu)\;\beta &=& 0 \\
( x_+^{N+1} -\mu )x_+\; \alpha + ( x_-^{N+1} -\mu )x_- \; \beta &=& 0
\end{eqnarray*}
The compatibility of this system yields the spectral curve
\begin{equation}
\mu + \mu^{-1}= x_+^{N+1}+x_-^{N+1} \equiv t(\lambda)
\label{spec0}
\end{equation}

We now impose that the curve passes through the point $(\lambda = 2, \mu = \mu_0)$
where $\mu_0$ is related to the value of the Casimir function by
\begin{equation}
K = \mu_0 + \mu_0^{-1}
\label{casimir}
\end{equation}
Setting
$$
x_\pm(2) = {1\over \sqrt{a}} \pm \sqrt{{1\over a} -1} = e^{\pm \alpha}, \quad 
\mu_0 = e^{ip_0}
$$
eq.(\ref{spec0}) gives  $\alpha = i{p_0\over N+1}$.
Hence the constant $a$ is related to the value of the  zero mode $p_0$ by:
$$
\sqrt{a} = {1\over \cos {p_0\over N+1}}
$$

The components of the eigenvector, properly normalized,  are meromorphic functions on the spectral curve
eq.(\ref{spec0}).  Choosing the normalization $\psi_{N+1}=\mu$, they read
\begin{equation}
\psi_i(\lambda,\mu) = {P_{N+1-i}(z) + \mu P_i(z) \over P_{N+1}(z)},\quad z={\lambda\over \sqrt{a}}
\label{psitrivial}
\end{equation}
We have introduced the polynomials of degree $j-1$:
\begin{equation}
P_j(z) = {x_+^j - x_-^j \over x_+ - x_- }, \quad P_j(z) = z^{j-1} + O(z^{j-3})
\label{tchebi}
\end{equation}
The first few polynomials are
$$
P_0=0,\quad P_1 = 1,\quad P_2=z,\quad P_3=z^2-1,\quad
P_4 = z^3-2z
$$
They are essentially the Tchebitchev polynomials of the second kind.

As we will see, eq.(\ref{psitrivial}) is the general form of the meromorphic function
$\psi_i(\lambda,\mu)$ even when  $a_i \neq a$. In particular, in order to take the continuum
limit, the poles of the eigenvector\footnote{In fact in this simple case the eigenvector has no  poles at finite distance because they are compensated by zeroes in the numerator.  This degeneracy is  lifted as soon as the $a_i$ are not all equal.} will have to be close to  the  roots of the equation
$P_{N+1}({\lambda \over \sqrt{a}}) = 0$, that is
\begin{equation}
\lambda_k^{(0)} = 2 \sqrt{a} \cos {Z_k^{(0)} \over N+1}, \quad Z_k^{(0)}= k\pi \quad k = 1,\cdots,N
\label{freelambda}
\end{equation}
For these special values  $\lambda_k^{(0)}$, we have
$x_\pm = e^{\pm i {k\pi \over N+1}} $ and eq.(\ref{spec0}) gives
$ \mu^{(0)}_k = (-1)^k $.
The set of points $(\lambda_k^{(0)}, \mu^{(0)}_k ), k = 1,\cdots,N,$ will be called the free configuration
and will play an important role below.

It is simple to take the continuum limit in this free case. We set 
$$
\lambda \simeq 2 - \Delta^2 \Lambda^2,\quad z \simeq 2 - \Delta^2 Z^2,
\quad \psi_{i\pm1} = \psi(x\pm \Delta),\quad
\Delta = {1\over N+1}
$$
where we have introduced the variable
\begin{equation}
Z = \sqrt{\Lambda^2 + p_0^2}
\label{defZ}
\end{equation}
The eigenvector equation becomes the Schroedinger equation
\begin{equation}
- \psi''(x) - p_0^2\psi(x) = \Lambda^2 \psi(x), \quad x = \Delta j
\label{freeschroe}
\end{equation}
We also have $x_\pm = 1 \pm i\Delta Z$
and the equation of the spectral curve reads:
$$
\mu + \mu^{-1} = (1+i\Delta Z)^{1/\Delta} + (1-i\Delta Z)^{-{1/ \Delta}}
$$
In the limit $\Delta \to 0$, it becomes
\begin{equation}
\mu + \mu^{-1} = 2 \cos Z
\label{freecurve}
\end{equation}
Similarly, the eigenvector becomes a  Baker-Akhiezer function:
\begin{equation}
\psi(x) = {\sin Z(1-x) + \mu \sin Z x \over \sin Z }
\label{freepsi}
\end{equation}
When $\mu = e^{\pm i Z}$ this reduces to $\psi(x) = e^{\pm  i Z x}$
 as it should be. Notice that when $\mu$ is kept as a free parameter, the above formula gives two independent solutions of eq.(\ref{freeschroe}), but when $\mu$ belongs to the  spectral curve eq.(\ref{freecurve}) one has
$$
\psi(x+1) = \mu \psi(x)
$$
Eq.(\ref{freepsi}) presents the two Bloch waves as a single function on the 
hyperelliptic spectral curve eq.(\ref{freecurve}).

\bigskip

Another example, important  to us,  will be the Dirac comb.
$$
[-\partial^2 - H_0 \delta (x) ] \psi (x) = \Lambda^2 \psi (x), \quad \delta(x+1) =  \delta(x)
$$
On each interval $x_j = j < x < x_{j+1} = j+1$, one has
$$
\psi (x) = \alpha_j e^{i\Lambda x} + \beta_j e^{-i\Lambda x} , \quad x_j < x <  x_{j+1}
$$
The Bloch condition 
$$
\psi (x + x_j) = \mu^j \psi (x),\quad 0 < x < 1
$$
gives $\alpha_j = (\mu e^{-i\Lambda })^j \alpha_0$, $ \beta_j = (\mu e^{i\Lambda })^j \beta_0$.
The continuity of $\psi(x)$ gives $\alpha_0 = - {(\mu - e^{-i\Lambda})  \over (\mu - e^{i\Lambda} )} \beta_0$, 
while the gap equation on the first derivative, $-\psi'(x_j+0) + \psi'(x_j-0) - H_0 \psi(x_j) = 0$,
gives the spectral curve
\begin{equation}
  \mu + \mu^{-1} = 2 \cos \Lambda - H_0 {\sin \Lambda \over \Lambda} 
  \label{speccomb}
\end{equation}
The Bloch wave itself is $\psi (x) = \mu^j \psi_{vac}(x-x_j)$,  $x_j < x < x_{ j+1}$,
where $\psi_{vac}(x)$ is given by eq.(\ref{freepsi}) (with $p_0=0$) but $\Lambda, \mu$ now belonging to the curve eq.(\ref{speccomb}).

\section{Separated variables.}

In this section, we  generalize the previous analysis when $a_i \neq a $
and express the dynamical variables of the Volterra model in terms 
of the separated variables.  Equivalent formulae where already obtained
a long time ago in \cite{KvM, vM}. 
A quantum version of this construction for the closed Toda chain can be found in \cite{Ba04}.

We have to reconstruct the eigenvectors  $\Psi$ of $L(\mu)$. Let us set
$\Psi = (\psi_i)$, $i=1, \cdots , N+1$. 
We normalize the last component $\psi_{N+1} = \mu$. 
Notice that due to eq.(\ref{gamma}), $\mu$ does not vanish  for finite $\lambda$. 
The components $\psi_i$ are meromorphic functions on the spectral curve
and are uniquely characterized by their poles and behavior at infinity which we now describe.

We will call $P^+(\lambda=\infty, \mu=\infty)$ and $P^-(\lambda=\infty, \mu=0)$ the two points above $\lambda = \infty$.  In the neighbourhood of $P^\pm$, the
local parameter is $\lambda^{-1}$ and  we have by direct  expansion
of eq.(\ref{gamma}):
\begin{eqnarray}
P^+ &:& \mu = {\cal A}^{-1} \lambda^{N+1}\Big(1  +
O(\lambda^{-2})\Big) \label{P+} \\
\quad P^- &:& \mu =  {\cal A}\lambda^{-N-1}\Big(1  +
O(\lambda^{-2})\Big) \label{P-}
\end{eqnarray}

At  the points $P^+$ and $P^-$, the
eigenvector $\Psi(P)$ behaves as:
\begin{eqnarray}
\psi_i (P)&=&{1\over \sqrt{a_{N+1} a_1 a_2 \cdots a_{i-1}}}\; \lambda^i \Big( 1  + O(\lambda^{-2}) \Big), 
\quad  P \sim P^+ \label{+infty} \\
\psi_i (P)&=& \sqrt{a_{N+1} a_1 a_2 \cdots a_{i-1}}\; \lambda^{-i} \Big( 1 
+ O(\lambda^{-2}) \Big), \quad  P \sim P^- \label{-infty}
\end{eqnarray}
This is easily deduced by inspection of eqs.(\ref{recur}).
 
From the general results of the classical inverse scattering theory,  we expect
$g+(N+1)-1=2N$ poles for the eigenvector (see e.g. \cite{DuKrNo90, BaBeTa03}).  From eq.(\ref{+infty}), 
we see that we have a fixed pole of order $N$ at $P^+$ (on the component $\psi_N$), and
there remains $g=N$ poles at finite
distance, the so called {\it dynamical poles}.  But we notice the 
symmetry property
$$
\psi_i(-\lambda,-\mu) = (-1)^i \psi_i(\lambda,\mu)
$$
so that the dynamical poles come in pairs
$$
\lambda_{N+1-k } = - \lambda_k,\quad \mu_{N+1-k} = - \mu_k
$$
and only
$(\lambda_k,\mu_k), k=1\cdots n,$ are independent parameters. 

Everything can be expressed in terms of these $2n = N$ quantities $(\lambda_k,\mu_k), k=1\cdots n$. In fact, they can be viewed as coordinates on (an open set of) phase space. 

First, the commuting Hamiltonians are easy to reconstruct. Indeed
the spectral curve is determined by requiring that it passes through
 the points $( \lambda_k,\mu_k), k=1\cdots n$,
 and through the point $(2,\mu_0)$, where $\mu_0$ is related to the Casimir
 function as in eq.(\ref{casimir}).

 The equation of the curve itself can be written as a determinant
 \begin{equation}
 \det \pmatrix{
  \lambda & \lambda^3 & \cdots & \lambda^{N+1} & \mu + \mu^{-1} \cr
       2       &        2^3    &  \cdots &       2^{N+1}     &  \mu_0 + \mu_0^{-1} \cr
 \lambda_1 & \lambda_1^3 & \cdots & \lambda_1^{N+1} & \mu_1 + \mu_1^{-1}   \cr
   \vdots      &                      &            &                              &              \vdots          \cr
 \lambda_n & \lambda_n^3 & \cdots & \lambda_n^{N+1} & \mu_n + \mu_n^{-1}
 }       =0
 \label{spectralcurve}
\end{equation}
Expanding over the first row, we obtain a curve of the form eq.(\ref{gamma}), and we can read directly
the Hamiltonians as the coefficients of $t(\lambda)$. They appear as functions of the $(\lambda_k,\mu_k)$ and can be shown to Poisson commute (see \cite{At68, EnRu01, BaTa} for a proof and for the quantum generalization of this fact).

Eqs.(\ref{+infty},\ref{-infty}) and the data of the $N$ dynamical poles also determine the functions $\psi_i$ uniquely. 
Being meromorphic functions
on a hyperelliptic curve, we can write quite generally 
\begin{equation}
\psi_i = { Q^{(i)}(\lambda) + \mu R^{(i)}(\lambda) \over \prod_{k=1}^n (\lambda^2 - \lambda_k^2) }
\label{psii1}
\end{equation}
where $Q^{(i)}$ and $R^{(i)}$ are polynomials such that
$$
 Q^{(i)}(-\lambda) = (-1)^i  Q^{(i)}(\lambda), \quad
  R^{(i)}(-\lambda) =(-1)^{i+1}  R^{(i)}(\lambda)
$$

Above $\lambda_k$, we have two points on the curve: $(\lambda_k, \mu_k)$ and
$(\lambda_k, \mu^{-1}_k)$. We want the poles to be at $(\lambda_k, \mu_k)$ only 
so that the numerator in eq.(\ref{psii1}) should vanish at the points  $(\lambda_k, \mu_k^{-1})$. This gives
$n$ conditions
\begin{equation}
Q^{(i)}(\lambda_k) + \mu_k^{-1} R^{(i)}(\lambda_k) = 0, \quad k = 1 \cdots n
\label{conditions}
\end{equation}
To have a pole of order $i$ at $P^+$ and a zero of order $i$ at $P^-$ we must choose
$$
{\rm degree~} Q^{(i)} = N-i, \quad {\rm degree~} R^{(i)} = i-1
$$
Hence, these two polynomials depend altogether on $n+1$ coefficients which are determined by imposing
the $n$ conditions eq.(\ref{conditions}) and requiring that the normalizations coefficients are inverse to each other at $P^\pm$ as in eqs.(\ref{+infty},\ref{-infty}). 

It is convenient to use the basis of polynomials $P_j(\lambda )$ given by eq.(\ref{tchebi}). 
We will write the formulae for $\psi_i$ in the case $i$ {\em odd}, the case
$i$ {\em even} is similar.

The polynomial $Q^{(i)}(\lambda )$ can be expanded over
$$
Q^{(i)}(\lambda ): \quad P_2(\lambda ),P_4(\lambda ),\cdots P_{N+1-i}
$$
and the polynomial $R^{(i)}(\lambda )$ can be expanded over
$$
R^{(i)}(\lambda ):\quad P_1(\lambda ),P_3(\lambda ),\cdots P_i(\lambda )
$$
Solving the linear system eq.(\ref{conditions}), the eigenvector can be written as
\begin{equation}
\psi_i = {K_i\over \prod(\lambda ^2 - \lambda _k^2)}\det \pmatrix{ 
\mu P_1(\lambda )  & \cdots & \mu P_i(\lambda )&
-P_{N+1-i}(\lambda ) &\cdots  &-P_2(\lambda ) \cr
 P_1(\lambda _1)  & \cdots &  P_i(\lambda _1)&
-\mu_1P_{N+1-i}(\lambda _1) &\cdots  &-\mu_1P_2(\lambda _1) \cr
\vdots & \vdots & \vdots &\vdots&\vdots& \vdots\cr
 P_1(\lambda _k)  & \cdots &  P_i(\lambda _k)&
-\mu_kP_{N+1-i}(\lambda _k) &\cdots  &-\mu_kP_2(\lambda _k) \cr
\vdots & \vdots & \vdots &\vdots&\vdots& \vdots\cr
 P_1(\lambda _n)  & \cdots &  P_i(\lambda _n)&
-\mu_nP_{N+1-i}(\lambda _n) &\cdots  &-\mu_nP_2(\lambda _n) 
}
\label{psii}
\end{equation}
where $K_i$ are constants independant of $\lambda,\mu$. Defining
\begin{equation}
\Theta_i =  \pmatrix{ 
 P_1(\lambda _1)  & \cdots &  P_{i-2}(\lambda _1)&
-\mu_1P_{N+1-i}(\lambda _1) &\cdots  &-\mu_1P_2(\lambda _1) \cr
\vdots & \vdots & \vdots &\vdots&\vdots& \vdots\cr
 P_1(\lambda _k)  & \cdots &  P_{i-2}(\lambda _k)&
-\mu_kP_{N+1-i}(\lambda _k) &\cdots  &-\mu_kP_2(\lambda _k) \cr
\vdots & \vdots & \vdots &\vdots&\vdots& \vdots\cr
 P_1(\lambda _n)  & \cdots &  P_{i-2}(\lambda _n)&
-\mu_nP_{N+1-i}(\lambda _n) &\cdots  &-\mu_nP_2(\lambda _n) 
}
\label{defthetai}
\end{equation}
we can compute the  leading terms in eq.(\ref{psii}) when $\lambda  \to \infty$. At 
$P^{(+)}$ the leading term comes from $\mu P_i(\lambda)$, while at $P^{(-)}$ it comes from $ P_{N+1-i}(\lambda)$.
\begin{eqnarray*}
\psi_i &\simeq& (-1)^{i-1\over 2} {\cal A}^{-1} K_i \det \Theta_i\; \lambda ^i,\quad P^+ \\
\psi_i &\simeq& (-1)^{i-1\over 2}K_i  \det \Theta_{i+2}\; \lambda ^{-i}, \quad P^- 
\end{eqnarray*}
Imposing that the two coefficients of $\lambda^i$ and $\lambda^{-i}$ are inverse to each other, we get
$$
K_i^2 = {{\cal A}  \over \det \Theta_i \det \Theta_{i+2}}
$$

Comparing with eqs.(\ref{+infty}, \ref{-infty}), we finally obtain
\begin{equation}
a_i =  {\det \Theta_{i} \det \Theta_{i+3} \over \det \Theta_{i+1}\det \Theta_{i+2}},\quad
a_N = {\det \Theta_{N}  \over \det \Theta_{N+2}}{\cal A},\quad
 a_{N+1}= {\det \Theta_3 \over \det \Theta_1} {\cal A}
 \label{aisep}
\end{equation}
Here ${\cal A}^{-1}$ is the coefficient of $\lambda^{N+1}$ in $t(\lambda)$, eq.(\ref{toflambda}),
 computed from eq.(\ref{spectralcurve}).

We impose the Poisson bracket on the variables $\lambda_k,\mu_k$ 
\begin{equation}
\{\lambda_k ,\lambda_{k'} \} = 0,\quad 
\{\lambda_k ,\mu_{k'} \} = - {1\over 2} \delta_{kk'}\left(4\lambda_k -  \lambda_k^3\right) \mu_k,\quad 
\{\mu_k ,\mu_{k'} \} = 0 
\label{poisep}
\end{equation}
One can then check that the Hamiltonians defined by eq.(\ref{spectralcurve}) are all 
in involution (this is a general result), and that the $a_i$ defined above do satisfy the 
Faddeev-Takhtadjan Poisson bracket. The fact that the expressions for $a_N, a_{N+1}$ are different 
from the ones in the bulk  is due to the choice of normalization of the eigenvector. However, the 
Poisson bracket of the $a_i$ is periodic. All this can be proved using techniques similar 
to the ones in \cite{Ba04}.

\section{Continuum limit of the spectral curve.}

We now take the continuum limit of the spectral curve eq.(\ref{spectralcurve}). The result is eq.(\ref{specurveH0Hm}).
 We set
$$
\lambda = \sqrt{a} z, \quad \lambda_k = \sqrt{a} z_k,\quad {2\over \sqrt{a} } = 2 \cos {p_0\over N+1} = z_0
$$
From these, the scaled variables $\Lambda, Z, Z_k$ are defined like this:
\begin{equation}
\lambda =  2 \cos {\Lambda \over N+1}, \quad z =  2 \cos {Z\over N+1}, \quad z_k = 2 \cos {Z_k  \over N+1}
\label{scaledvariables}
\end{equation}
Notice that we have $Z = \sqrt{\Lambda^2 + p_0^2}$.
In the following, we will refer to the terminology "perturbation theory" when  the points $(Z_k, \mu_k)$ are small deviations from the free configuration eq.(\ref{freelambda}). 
The formulae we will write will make sense  
in this perturbative setting. This however does not exclude the possibility to have a finite number of points which are large deviation.
We will also be interested in the deviation from the zero mode configuration. That is 
we make the substitution $\sqrt{a_i} \to \sqrt{a} \sqrt{\tilde{a}_i}$ everywhere on the lattice. Alternatively
this amounts to using the variable $ z = \lambda / \sqrt{a}$.

Using the basis of polynomials $P_j(z)$ defined in eq.(\ref{tchebi})  instead of the
$z^j$, we can write the spectral curve as (it has the right form and passes through the right points)
 $$
 \det \pmatrix{
 \mu + \mu^{-1} &  P_{N+2}(z) &  \cdots & P_4(z) & P_{2}(z)  \cr
 \mu_0 + \mu_0^{-1} &  P_{N+2}(z_0) &   \cdots &       P_4(z_0)    &    P_2(z_0)  \cr
 \mu_1 + \mu_1^{-1} &  P_{N+2}(z_1)  &\cdots &   P_4(z_1) & P_2(z_1)   \cr
   \vdots      &                      &            &                              &              \vdots          \cr
 \mu_n + \mu_n^{-1} &P_{N+2}(z_n)& \cdots &P_4(z_n) & P_2(z_n) 
 }       =0
$$
Without changing the determinant, we can subtract to the first column the linear combination
of the next two columns:
$$
P_{N+2}(z_k) - P_N(z_k) = 2 \cos Z_k
$$
The first column becomes
$$
\pmatrix{  \mu + \mu^{-1}  - 2 \cos Z \cr
\gamma_0 \cr
\gamma_1 \cr 
\vdots \cr
\gamma_n}
$$
where we have set
\begin{equation}
\gamma_k = \mu_k + \mu_k^{-1}  - 2 \cos Z_k
\label{gammak}
\end{equation}
Notice that $\gamma_0 = 0$. The reason for this subtraction is that
for the free configuration we also have $\gamma_k^{(0)}=0$, so that  the spectral curve becomes simply
$ \mu + \mu^{-1}  = 2 \cos Z$
as it should be. The subtraction gives sense to the spectral curve in perturbation theory.
Expanding the determinant over the first row, we can write
$$
\mu + \mu^{-1} - 2 \cos Z = \sum_{j=1}^{n+1} H_{2j} P_{2j}(z)
$$
The $H_{2j}$ are given by
$$
H = N^{-1} V
$$
where we have defined
$$
H = \pmatrix{H_{N+2} \cr H_{N} \cr \vdots \cr H_2},\quad
N = \pmatrix{
  P_{N+2}(z_0) &   \cdots &       P_4(z_0)    &    P_2(z_0)  \cr
  P_{N+2}(z_1)  &\cdots &   P_4(z_1) & P_2(z_1)   \cr
    \vdots           &            &                   &   \vdots      \cr
P_{N+2}(z_n)& \cdots &P_4(z_n) & P_2(z_n) 
 }
, \quad
V =\pmatrix{\gamma_0 \cr \gamma_1 \cr \vdots \cr \gamma_n}
$$
We will need to treat separately  the first row and column in the matrix $N$. Let us write it as
$$
N = \pmatrix{ A & B \cr C & D}
$$
where
$$
A = P_{N+2}(z_0),\quad B = \pmatrix{ P_N(z_0) & \cdots P_4(z_0) & P_2(z_0) }
$$

$$
C = \pmatrix{ P_{N+2}(z_1) \cr \vdots \cr P_{N+2}(z_n)},\quad
D = \pmatrix{   P_{N}(z_1)  &\cdots &   P_4(z_1) & P_2(z_1)   \cr
    \vdots           &            &                   &   \vdots      \cr
P_{N}(z_n)& \cdots &P_4(z_n) & P_2(z_n) 
 }
$$

To zero-th order in perturbation theory, we denote $N = N^{(0)} $ and similarly $A^{(0)}, B^{(0)}, C^{(0)},  D^{(0)} $. 
To take the continuum limit we have to consider the matrix $N N^{(0)-1}$.

\begin{lemma}
In the continuum limit, we have
 \begin{equation}
 NN^{(0)-1} = \pmatrix{ 1 & 0 \cr 
 {\sin Z_k \over Z_k }  {p_0 \over\sin p_0 } &
   {\sin Z_k \over Z_k }  
   \left\{ {1\over Z_k^2 - m^2\pi^2} -  {1\over p_0^2 - m^2\pi^2} \right\} 2(-1)^{m} m^2\pi^2
   }, \quad k,m = 1, \cdots , \infty
 \label{NN0-1}
\end{equation}
\end{lemma}
\proof
Since  $P_{N+2}(z) = z P_{N+1}(z) - P_N(z)$
and $P_{N+1}(z_k^{(0)}) = 0$, 
we have
$$
 C_k^{(0)} = - D_{k1}^{(0)} \Longrightarrow 
(D^{(0)-1} C^{(0)} )_k = - \delta_{k1}
$$
so that
$$
N^{(0)-1} ={1\over A + B_1} \pmatrix{ 1 & -BD^{(0)-1} \cr
F & (A+B_1)D^{(0)-1} - F \otimes
BD^{(0)-1}
}
$$
where $F$ is the column vector with components $F_k = \delta_{k,1}$, $k=1,\cdots, n$.
It follows that
$$
NN^{(0)-1} = 
\pmatrix{1 & 0 \cr {1\over A+B_1}(C+DF) & D D^{(0)-1} -{1\over A+B_1}(C+D F) \otimes BD^{(0)-1} }
$$
Noticing that
$$
A+B_1 = P_{N+2}(z_0) + P_N(z_0)
$$
$$
(C+DF)_k = P_{N+2}(z_k) + P_N(z_k) 
$$
we get  in the continuum limit 
$$
 {1\over A+B_1}(C+DF)_k 
 \to {\sin Z_k \over Z_k }{ p_0 \over \sin p_0}
 $$
 
The  main trick  to proceed is an explicit formula for the inverse of the matrix $D^{(0)}$. It is not difficult to check that
$$
(D^{(0)-1})_{jk} =  {4 \over N+1} \sin^2{k\pi\over N+1} \; P_{2j}(z_k^{(0)})
$$
With this,  we find  
 \begin{eqnarray*}
  (B D^{(0)-1})_m &=& {4\over N+1} \sin^2 {m\pi \over N+1} \sum_{j=1}^n P_{2j}(z_0) P_{2j}(z_m^{(0)}) 
  \\ 
  &=&   { \sin^2 {m\pi \over N+1} \over \sin {p_0\over N+1} \sin {m\pi\over N+1}} 
 {4\over N+1} \sum_{j=1}^n \sin {2j p_0\over N+1} \sin {2j m\pi \over N+1} 
 \to 2 {m\pi \over p_0} \int_0^1 dx \sin p_0 x \sin m\pi x 
 \end{eqnarray*}
 and the last integral is easily evaluated with the result
  $$
 (B D^{(0)-1})_m \to  (-1)^m {2  m^2 \pi^2\sin p_0 \over p_0 (p_0^2 - m^2\pi^2)}
 $$
 Similarly we compute
 $$
 (DD^{(0)-1})_{km} = {\sin Z_k \over Z_k } {
 (-1)^{m} 2 m^2\pi^2 \over Z_k^2 - m^2\pi^2}
 $$
 Gathering  all this we get eq.(\ref{NN0-1}).
 \square
 
 \bigskip
 
We now introduce the important infinite matrix
\begin{equation}
M_{km} = {1\over Z_k^2 - m^2\pi^2}, \quad k,m = 1, \cdots , \infty
\label{defM}
\end{equation}
and the important vector $\vert \eta \rangle$
\begin{equation}
\vert \eta \rangle = M^{-1} \pmatrix{1 \cr 1\cr  \vdots \cr 1}
,\label{defetam}
\end{equation}

With these notations we can compute the inverse of the matrix $NN^{(0)-1}$:

\begin{lemma}
$$
(NN^{(0)-1})^{-1} =\pmatrix{ 1 & 0 \cr {p_0\over \sin p_0}\left( {1  \over 1 - \langle \chi(p_0) \vert \eta \rangle}  \right) {(-1)^{m+1} \over 2 m^2\pi^2} \eta_m  &
{(-1)^{m} \over 2 m^2 \pi^2} \left \{ \left( 1 +{ \vert \eta \rangle\langle \chi (p_0) \vert  \over 1 - \langle \chi (p_0)\vert \eta \rangle} 
 \right)M^{-1}\right\}_{mk}  {Z_{k} \over \sin Z_{k} }}
$$
where we have defined the vector $\langle \chi(p_0) \vert _m = {1\over p_0^2 - \pi^2 m^2}$.
\end{lemma}
\proof
With the above notations we can write
  $$
 NN^{(0)-1} = \pmatrix{ 1 & 0 \cr 
 {\sin Z_k \over Z_k }  {p_0 \over\sin p_0 } &
   {\sin Z_k \over Z_k }  M_{kl} \left( \delta_{lm} - \eta_l \chi_{m}(p_0) \right)
   2(-1)^{m} m^2\pi^2
   }
 $$
Letting
$$
NN^{(0)-1} = \pmatrix{1 & 0 \cr Y & X } \Longrightarrow
(NN^{(0)-1})^{-1}= 
\pmatrix{ 1 & 0 \cr -X^{-1}Y & X^{-1} }
$$
we find
\begin{equation}
(X^{-1})_{mk} = {(-1)^{m} \over 2 m^2 \pi^2} \left \{ \left( 1 +{ \vert \eta \rangle\langle \chi (p_0) \vert  \over 1 - \langle \chi (p_0)\vert \eta \rangle} 
 \right)M^{-1}\right\}_{mk}  {Z_{k} \over \sin Z_{k} }
 \label{Xinv}
\end{equation}
and
$$
( -X^{-1}Y)_m = {p_0\over \sin p_0}\left( {1  \over 1 - \langle \chi(p_0) \vert \eta \rangle} 
 \right) {(-1)^{m+1} \over 2 m^2\pi^2} \eta_m 
$$ 
\square

Let us return to the formula for the spectral curve. We assume that the conditions explained in Section 9 are satisfied so that the infinite sums we will manipulate are convergent. Denote
\begin{equation}
\eta(Z) = 1 - \sum_{m=1}^\infty {\eta_m \over Z^2 - m^2\pi^2}
\label{etaZ}
\end{equation}
and
\begin{equation}
\vert \Gamma \rangle_m = \sum_{k} M^{-1}_{mk} {Z_{k} \over \sin Z_{k} }\gamma_{k}
\label{defGammam}
\end{equation}
These quantities enter the expression of the continuum limit of the spectral curve. 
\begin{proposition}
In the continuum limit, the equation of the spectral curve becomes:
\begin{equation}
\mu + \mu^{-1} = 2 \cos Z + {\sin Z \over Z} \left( -H_0 + \sum_m {H_m \over Z^2 - m^2\pi^2} \right)
\label{specurveH0Hm}
\end{equation}
where the conserved quantities $H_m$ can be taken as
\begin{equation}
 H_m =  \Gamma_m + {1\over \eta(p_0)}  \langle \chi(p_0) \vert \Gamma  \rangle   \eta_m,
 \quad H_0 = \sum_m {H_m \over p_0^2 - \pi^2 m^2} = {1\over \eta(p_0)}\sum_m {\Gamma_m \over p_0^2 - \pi^2 m^2} 
 \label{HmH0}
\end{equation}
\end{proposition}
\proof
We have
$$
\mu + \mu^{-1} - 2 \cos Z= \sum_{i=0}^n P_{N+2-2i}(z) H_{N+2-2i}  = \sum_{i=0}^n P_{N+2-2i}(z)
(N^{-1}V)_i
$$
we  insert $1 =  N^{(0)-1} N^{(0)}$ into the above expression:
\begin{equation}
\mu + \mu^{-1} - 2 \cos Z= \sum_{i,j,k} P_{N+2-2i}(z)( N^{(0)-1})_{ik} ( N^{(0)})_{kj}(N^{-1}V)_j
\label{speccurve2}
\end{equation}
Hence, we need to compute 
 \begin{eqnarray*}
 \sum_{i} P_{N+2-2i}(z)( N^{(0)-1})_{im} &=& \\ 
 && \hskip -4cm \left( {P_{N+2}(z) + P_N(z) \over P_{N+2}(z_0) + P_N(z_0) },
 - {P_{N+2}(z) + P_N(z) \over P_{N+2}(z_0) + P_N(z_0) } BD^{(0)-1} +
  \sum_{i} P_{N+2-2i}(z)( D^{(0)-1})_{im} \right)
\end{eqnarray*}
whose limit $N \to \infty$ is easy to take
 \begin{eqnarray*}
 \sum_{i} P_{N+2-2i}(z)( N^{(0)-1})_{im} &\to&  
{\sin Z \over Z }  \left(  { p_0 \over \sin p_0 },
 2(-1)^m m^2\pi^2  \left\{  {1 \over Z^2 - m^2\pi^2 } - {1 \over p_0^2 - m^2\pi^2 } \right\} \right)
\end{eqnarray*}
We can now  take the limit $N\to \infty$ in eq.(\ref{speccurve2})
 \begin{eqnarray*}
\mu + \mu^{-1} - 2 \cos Z&=&  
{\sin Z \over Z }  \left(  { p_0 \over \sin p_0 }\widetilde{H}_0 + \sum_{m=1}^\infty
 2(-1)^m m^2\pi^2  \left\{  {1 \over Z^2 - m^2\pi^2 } - {1 \over p_0^2 - m^2\pi^2 }  \right\} 
 \widetilde{H}_m\right)
\end{eqnarray*}
where
$$
\widetilde{H}_m = (NN^{(0)-1})^{-1} \pmatrix{\gamma_0 \cr 
\gamma_1 \cr \vdots \cr \gamma_n} = \pmatrix{0 \cr X^{-1} \pmatrix{\gamma_1 \cr
\vdots \cr \gamma_n }  }
$$
where $X^{-1}$ is given in eq.(\ref{Xinv}) and we  remembered that $\gamma_0=0$.
Since  $1 - \langle \chi (p_0) \vert \eta \rangle = \eta(p_0)$  the equation of the spectral curve finally becomes
\begin{eqnarray*}
\mu + \mu^{-1} -2 \cos Z &=& {\sin Z \over Z} \left(   \sum_m \left( {1\over Z^2 - m^2\pi^2}-  {1\over p_0^2 - m^2\pi^2}\right)
\left( \Gamma_m + {1\over \eta(p_0)} \eta_m \langle \chi(p_0) \vert \Gamma  \rangle 
 \right)
\right)
\end{eqnarray*}
\square

Another useful expression of this result is:
\begin{equation}
\mu + \mu^{-1} = 2 \cos Z + {\sin Z \over Z} \left(\sum_m {\Gamma_m \over Z^2 - m^2\pi^2}
- { \eta(Z)\over \eta(p_0)}  {\Gamma_m \over p_0^2 - m^2\pi^2} \right)
\label{specurveGammaEta}
\end{equation}

The next proposition performs a few consistency checks.

\begin{proposition}
The points $(Z= p_0, \mu_0^{\pm 1})$, and $(Z = Z_k, \mu_k^{\pm 1})$, all belong to the curve eq.(\ref{specurveGammaEta}).
\end{proposition}
\proof
When $Z = p_0$,  we  find  $\mu + \mu^{-1} = 2 \cos p_0$, hence the  curve passes through the point $\Lambda = 0, \mu_0^{\pm 1}$, as it should be.
When $Z = Z_k$, recalling that $\eta(Z_k) = 0$, we find
\begin{eqnarray*}
\mu + \mu^{-1} &=&2 \cos Z_k +  {\sin Z_k \over Z_k} \sum_{m}
{1\over Z_k^2 - m^2\pi^2} M^{-1}_{m l} {Z_l \over \sin Z_l} (\mu_l + \mu_l^{-1} -2 \cos Z_l )\\
&=& 2 \cos Z_k +   {\sin Z_k \over Z_k} \sum_{m}
M_{km} M^{-1}_{m l} {Z_l \over \sin Z_l} (\mu_l + \mu_l^{-1}-2\cos Z_l )=  \mu_k + \mu_k^{-1}
\end{eqnarray*}
Hence the curve passes through the points $Z_k, \mu_k^{\pm 1}$.
\square

We now show that the $H_m$ in eq.(\ref{specurveH0Hm})  all Poisson commute. We need the following result:
\begin{lemma}
One has
\begin{eqnarray}
\{ \Gamma_n, \Gamma_{m} \} &=& 0 \label{rel1}\\
\{ \Gamma_n , \eta_m \} &=& \{ \Gamma_m , \eta_n \}  \label{rel2} \\
\{ \eta_n, \eta_m \} &=& 0 \label{rel3}
\end{eqnarray}
\end{lemma}
\proof
Recall the definitions eqs.(\ref{defetam},\ref{defGammam}) of $\eta_m$ and $\Gamma_m$.
$$
\eta_m = M^{-1}_{mk} \vert 1 \rangle_k, \quad \Gamma_m = M^{-1}_{mk} \vert \tilde{\gamma} \rangle_k,
\quad  \tilde{\gamma}_k = {Z_k \over \sin Z_k} \gamma_k
$$
where $M^{-1}_{mk}$ is the inverse of the matrix defined  in eq.(\ref{defM}).
The relation eq.(\ref{rel3}) is obvious because the $\eta_m$ depend only on the $Z_k$. Consider the second relation eq.(\ref{rel2}): 
\begin{eqnarray*}
\{ \eta_n , \Gamma_m  \} &=& \{ M^{-1}_{nk} 1_k , M^{-1}_{ml}\tilde{\gamma}_l \} 
= M^{-1}_{m,l} \{ M^{-1}_{nk} , \tilde{\gamma}_l \} 1_k \\
&=& -M^{-1}_{m,l}M^{-1}_{n,k'} \{ M_{k' p}, \tilde{\gamma}_l \} M^{-1}_{pk}  \vert 1 \rangle_k 
= - M^{-1}_{m,l}M^{-1}_{n,l} \{ M_{l p}, \tilde{\gamma}_l \} \eta_p
\end{eqnarray*}
where in the last step we used that $\{ M_{k' p}, \tilde{\gamma}_l \}  = 0$ if $k' \neq l$. The result is obviously symmetric in $m$ and $n$. Finally the first statement, eq.(\ref{rel1}), is simple.  One has
\begin{eqnarray*}
\{ \Gamma_m , \Gamma_n \} &=&  \{ M^{-1}_{mk} \tilde{\gamma}_k,  M^{-1}_{nl} \tilde{\gamma}_l \} =
 - M^{-1}_{mr} M^{-1}_{nl} \Big[ \{ M_{rs} , \tilde{\gamma}_l \} - \{  M_{ls}, \tilde{\gamma}_r  \} \Big] M^{-1}_{sk} \tilde{\gamma}_k 
\end{eqnarray*}
but because of the structure of $M$, we have $\{ M_{rs} , \tilde{\gamma}_l \} = 0$ if $r \neq l$ and for 
$ r = l$ the term in the square bracket obviously vanishes. This is a special case of a general theorem
\cite{EnRu01, BaTa}.
\square

We are now ready to prove
\begin{proposition}
The quantities $H_0$, $H_m$, Poisson commute
$$ 
\{ H_0, H_n \} = 0, \quad \{ H_n, H_{m}\} = 0
$$
\end{proposition}
\proof
Using eq.(\ref{HmH0}), one has
$$
\{ H_n, H_{m} \} = \eta_n ( \{ C, \Gamma_m \} + C \{ C , \eta_m \} ) -  \eta_m ( \{ C, \Gamma_n \} + C \{ C , \eta_n \} ) 
$$
where we denoted 
$$
C = {1\over \eta(p_0)}  \langle \chi(p_0) \vert \Gamma  \rangle = {1\over \eta(p_0)} \sum_l {\Gamma_l \over p_0^2 - \pi^2 l^2}
$$
One has
$$
\{ \Gamma_m, C \} = {1\over \eta(p_0)} C \sum_l { \{\Gamma_m, \eta_l \} \over p_0^2 - \pi^2 l^2},
\quad
\{ \eta_m, C \}  = {1\over \eta(p_0)}  \sum_l  { \{\eta_m, \Gamma_l \} \over p_0^2 - \pi^2 l^2}
$$
hence
$$
 \{ C, \Gamma_m \} + C \{ C , \eta_m \} = - {1\over \eta(p_0)} C \sum_l 
  {\{\Gamma_m, \eta_l \} +  \{\eta_m, \Gamma_l \} \over p_0^2 - \pi^2 l^2} = 0
 $$
\square

All this means that $(Z_k,\mu_k)$ are separated coordinates for the Hamiltonians $H_m$.

\section{Continuum limit of the eigenvector.}

Having found the continuum limit of the spectral curve, we now consider the 
limit of the eigenvector. Again, the continuum limit can be computed,  the result being
eq.(\ref{baker1}).

As seen from eq.(\ref{psii}), the eigenvector can be written as (for $i$ {\em odd})
$$
\psi_i = { \sqrt{\cal A}\over \prod(z ^2 - z _j^2)}{\det N_i  \over \sqrt{ \det \Theta_i \det \Theta_{i+2}}}
$$
where
\begin{equation}
N_i = \pmatrix{ 
\mu P_i(z )+P_{N+1-i}(z ) & \mu P_1(z )  & \cdots & 
-P_{N+1-i}(z ) &\cdots  &-P_2(z ) \cr
 P_i(z _1)+\mu_1P_{N+1-i}(z _1)&  P_1(z _1)  & \cdots & 
-\mu_1P_{N+1-i}(z _1) &\cdots  &-\mu_1P_2(z _1) \cr
\vdots & \vdots & \vdots &\vdots&\vdots& \vdots\cr
   P_i(z _k)+\mu_jP_{N+1-i}(z _k)& P_1(z _k)  & \cdots &
-\mu_kP_{N+1-i}(z _k) &\cdots  &-\mu_kP_2(z _k) \cr
\vdots & \vdots & \vdots &\vdots&\vdots& \vdots\cr
P_i(z _n)+\mu_nP_{N+1-i}(z _n)&  P_1(z _n)  & \cdots &  
-\mu_nP_{N+1-i}(z _n) &\cdots  &-\mu_nP_2(z _n) 
}
\label{ni}
\end{equation}
Compared to eq.(\ref{psii}), we have subtracted the $i$-th column to the first one for the same reason 
than in the previous section. Also we have used the variable $z$, $z_k$ instead of $\lambda$, $\lambda_k$.
Let us decompose the matrix $N_i$ in blocs  particularizing the first row and first column:
$$
N_i = \pmatrix{ U_i & V_i \cr W_i & \Theta_i }
$$
where
$$
U_i \equiv \mu P_i(z ) + P_{N+1-i}(z )
$$
$$
(W_i )_k \equiv \mu_k P_i(z _k) + P_{N+1-i}(z _k)
$$
$$
(V_i)_j= \mu P_j(z ) \theta(i-j) - P_{N+1-j}(z ) \theta(j-i),\quad i,j~~ odd
$$

\bigskip

To order zero in perturbation, we have $- (-1)^k P_{j}(z ^{(0)}_k) = P_{N+1- j}(z ^{(0)}_k)$ so that
$$
N_i^{(0)} =  \pmatrix{ U_i & V_i \cr 0 & \Theta_i^{(0)}}
$$
where $ \Theta_i^{(0)}$ is the matrix eq.(\ref{defthetai}) evaluated on the free configuration. It
is in fact  independent of $i$ and we will denote it by $ \Theta^{(0)}$. The appearance 
of zero in the lower left corner was the reason for the subtraction in eq.(\ref{ni}) and makes things better behaved
in perturbation.

The matrix $ N_i^{(0)}$ being bloc triangular we can compute its inverse:
$$
 N_i^{(0)-1} = \pmatrix{U_i^{-1} & - U_i^{-1} V_i \Theta^{(0)-1} \cr  0 &  \Theta^{(0)-1} }
$$
so that
$$
N_i N_i^{(0)-1} = \pmatrix{ 1 & 0 \cr U_i^{-1} W_i & \Theta_i \Theta^{(0)-1}-U_i^{-1} W_i\otimes  V_i \Theta^{(0)-1} }
$$
Returning to the formula for $\psi_i$, we multiply 
all the matrices by $\Theta^{(0)-1}$.  The factors  $\det \Theta^{(0)-1}$ cancel 
between the numerator and denominator.
We arrive at
$$
\psi_i = {\sqrt{\cal A} \;\;U_i \over \prod_k  z ^2 - z _k^2}\;\;
  {\det \Big(\Theta_i \Theta^{(0)-1}-U_i^{-1} W_i \otimes V_i \Theta^{(0)-1}\Big) \over 
\sqrt{\det \Big(\Theta_i \Theta^{(0)-1}\Big) \det \Big(\Theta_{i+2} \Theta^{(0)-1}\Big)}}
$$

We want to take the scaling limit of this expression. Again, the main trick is an explicit 
formula for the inverse of $\Theta^{(0)}$.
It is not difficult to check that

$$
(\Theta^{(0)-1})_{jk} =  {4 \over N+1} \sin^2{k\pi\over N+1} \; P_{j}(z _k^{(0)})
$$

Let us compute $\Theta_i\Theta^{(0)-1}$. Using the parametrization  eq.(\ref{scaledvariables})
we find (recall that $i$ is assumed to be odd)
\begin{eqnarray*}
(\Theta_i \Theta^{(0)-1})_{km}&=&{4\over N+1}{ \sin {m\pi\over N+1}\over \sin {Z_k \over N+1}} 
\left\{
\sum_{j=1, {\rm odd}}^{i- 2} \sin{jZ_k  \over N+1} \sin {jm\pi \over N+1}
\right. \\
&& \hskip 3cm  
\left. -  \mu_k 
\sum_{j=i, {\rm odd}}^{N-1} \sin{(N+1-j)Z_k  \over N+1} \sin {jm\pi \over N+1}
\right\}
\end{eqnarray*}
Defining $\Theta(x)$ as the scaling  limit of $\Theta_i \Theta^{(0)-1}$,
we find (there is a factor $1/2$ because the sum 
is over $j$ odd only)
\begin{eqnarray*}
({\Theta}(x))_{km} &=& 2{ m\pi \over Z_k } \left\{
\int_0^x dy \sin Z_k y \sin  m\pi y  -  \mu_k 
\int_x^1 dy  \sin Z_k(1-y) \sin m\pi y
\right\}
\end{eqnarray*}

Similarly, we define $U(x,\Lambda ,\mu)$ and $W_k(x)$ by
$$
U_i = (N+1) U(x,\Lambda ,\mu), \quad (W_i)_k = (N+1) {W}_k(x) 
$$
We find
$$
U(x,\Lambda ,\mu) = \mu {\sin x  Z  \over Z  }
+  {\sin (1- x)  Z  \over  Z  },\quad Z = \sqrt{\Lambda^2 + p_0^2}
$$
and
$$
W_k(x) =  {\sin Z_k x \over Z_k } 
+ \mu_k {\sin Z_k(1- x) \over Z_k }
$$

Finally, we have (again there is a factor $1/2$ because the sum 
is over $j$ odd only)
$$
(V_i \Theta^{(0)-1})_{m} = {V}_{m}(x,Z,\mu)
$$
where
\begin{eqnarray}
{V}_{m}(x,\Lambda ,\mu)=  2 m\pi \left\{  \int_0^x dy 
 \; \mu {\sin y Z \over Z }\sin m\pi y
- \int_x^1 {\sin (1- y) Z \over Z }\sin m\pi y
  \right\}
 \label{vkx}
\end{eqnarray}

Putting everything together, we arrive at (up to a factor\footnote{ In this factor we will include in particular $ {1 \over \prod_k  (1- Z_k^2/Z^2 )}$  
which produces  the poles at $Z^2 = Z_k^2$.  This is important for the analyticity properties 
of $\psi(x)$ but plays little role for the considerations of this  paper.}  independent of $x$
)

\begin{proposition}
\begin{equation}
\psi (x,\Lambda ,\mu) =  U(x,\Lambda ,\mu) -   \langle {V}(x,\Lambda ,\mu)\vert
 {\Theta}^{-1}(x) \vert {W}(x) \rangle 
 \label{baker1}
\end{equation}
where we denoted by $ \langle {V}(x,\Lambda ,\mu)\vert$ the row vector with components $V_m(x,\Lambda ,\mu)$
and by $\vert {W}(x) \rangle$ the column vector with components $W_k(x)$.
\end{proposition}
It is easy to show that the infinite sums involved in this formula converge under the conditions 
eq.(\ref{existencecondition}) of Section 9.

Eq.(\ref{baker1}) is the generalization of eq.(\ref{freepsi}). Here, the $\Lambda $ and $\mu$ dependence is entirely contained in the function $U(x,\Lambda ,\mu)$ and the vector $V(x,\Lambda ,\mu)$. For the moment 
they are free complex parameters.

\bigskip

We now want to specialize to  $\Lambda = 0,\mu_0 = e^{\pm i p_0}$.
We have
$$
U(x,\Lambda ,\mu)\vert_{0, e^{\pm i p_0} }=  {\sin p_0\over p_0 } e^{\pm ip_0 x}, \quad
V_{m}(x,\Lambda ,\mu)\vert_{0, e^{\pm i p_0} } =    U(x,\Lambda ,\mu)\vert_{0, e^{\pm i p_0} }
  \widetilde{V}_{m}^{(\pm)}(x,p_0)
$$
where
\begin{eqnarray*}
 \widetilde{V}_{m}^{(\pm)}(x,p_0) =   -m\pi 
 \left[ {e^{im\pi x }\over m\pi \pm p_0} +  {e^{-im\pi x }\over m\pi \mp p_0} \right]
 \end{eqnarray*}
 
 Hence, up to a constant
$$
\psi^{(\pm)}(x,p_0) =  e^{\pm ip_0 x}\Big[ 1 - 
\langle \widetilde{V}^{(\pm)}(x,p_0) \vert \Theta^{-1}(x) \vert W(x)\rangle  \Big]
$$
These are the primary fields of CFT. Their logarithmic derivatives are the 
free fields of the Coulomb gaz representation. Notice that we have two 
such fields playing a completely symmetrical role: we go from one to the other
by changing $p_0 \to - p_0$. This circumstance was recognized and used with 
great profit in \cite{GeNe}. The separated variables make this symmetry explicit
and built in.

\section{Schroedinger equation.}

Having found a formula for the wave function $\psi(x, \Lambda,\mu)$, the next
question is to find the potential in the Schroedinger equation that $\psi(x,\Lambda,\mu)$
 is expected to
satisfy. At this point it is simpler to forget the lattice model and 
work directly with  eq.(\ref{baker1}).

Let us denote by $\vert E(x) \rangle$ the vector with components
$$
E_m(x) = 2m\pi \sin m\pi x
$$
Calculating explicitly the integrals in eq.(\ref{vkx}), we find (' denotes the derivative with respect to $x$)
\begin{equation}
 {V}_{m}(x, \Lambda,\mu) = {U(x, \Lambda,\mu) E'_m(x) - U'(x, \Lambda,\mu) E_m(x) \over Z^2  - m^2\pi^2 },\quad Z^2 = \Lambda^2 + p_0^2
 \label{vkxbis}
 \end{equation}
 and similarly  for $\Theta(x)$, we find
 \begin{equation}
\Theta_{km}(x) = {W_k(x) E'_{m}(x) - W'_k(x) E_{m}(x) \over Z_k^2 - m^2 \pi^2 }
\label{thetax}
\end{equation}
Notice the important formulae 
 $$
 \langle V'(x, \Lambda,\mu)\vert = U(x, \Lambda,\mu)  \langle E(x)\vert, \quad
  \Theta'(x) =  \vert {W}(x) \rangle \langle E(x)\vert
$$
The derivative of the matrix $\Theta(x)$ is a rank one projector. The matrix $\Theta(x)$
has a form familiar in the theory of integrable systems, and we know that it leads to 
non linear differential equations. 
Indeed, let us define the vector $\vert \Phi(x) \rangle$ by
\begin{equation}
\vert \Phi (x) \rangle = \Theta^{-1}(x)  \vert W(x) \rangle 
\label{defPhi}
\end{equation}
and let $K$ be the diagonal matrix
$$
K_{mm'} = m\pi \; \delta_{mm'}
$$

\begin{proposition}
The vector $\vert \Phi(x) \rangle$ satisfies the set of coupled non linear second order 
differential equations
\begin{equation}
\vert \Phi'' \rangle + 2 (\langle \Phi \vert E \rangle)' \vert \Phi \rangle + K^2 \vert \Phi \rangle =0
\label{diffphi}
\end{equation}
where
$$
\langle \Phi \vert E \rangle (x) = \sum_m \Phi_m(x) E_m(x)
$$
\end{proposition}
\proof
 By derivation, and using the formula for $\Theta'$,
 we get
\begin{eqnarray*}
\vert W \rangle &=& \Theta \vert \Phi \rangle \\
\vert W' \rangle &=& \Theta' \vert \Phi \rangle + \Theta \vert \Phi' \rangle = \langle \Phi \vert E \rangle  
\vert W \rangle + \Theta \vert \Phi' \rangle  \\
\vert W'' \rangle &=&   \langle \Phi \vert E \rangle \vert W' \rangle
+ \Theta \vert \Phi'' \rangle +( \langle \Phi \vert E \rangle' + \langle \Phi' \vert E \rangle) \vert W \rangle
\end{eqnarray*}
But we also have
\begin{equation}
{\cal Z}^2 \Theta - \Theta K^2 = \vert W \rangle \langle E'\vert - \vert W' \rangle \langle E \vert
\label{eqint}
\end{equation}
where we defined the diagonal matrix ${\cal Z}$
$$
{\cal Z}_{kk'} = Z_k \delta_{kk'}
$$
Applying this identity to $\vert \Phi \rangle$, we get
$$
{\cal Z}^2 \Theta \vert \Phi \rangle - \Theta K^2 \vert \Phi \rangle =  \langle E' \vert \Phi \rangle \vert W \rangle
-\langle E \vert \Phi \rangle \vert W' \rangle
$$
But 
$$
{\cal Z}^2 \Theta \vert \Phi \rangle = {\cal Z}^2  \vert W \rangle = -  \vert W'' \rangle
$$
and therefore
$$
 \vert W'' \rangle = \langle E \vert \Phi \rangle \vert W' \rangle -  \langle E' \vert \Phi \rangle \vert W \rangle - \Theta K^2 \vert \Phi \rangle
$$
Comparing these two expressions of $ \vert W'' \rangle$, we find
$$
 \Theta \vert \Phi'' \rangle +( \langle \Phi \vert E \rangle' + \langle \Phi' \vert E \rangle) \vert W \rangle
 =  -  \langle E' \vert \Phi \rangle \vert W \rangle - \Theta K^2 \vert \Phi \rangle
$$
Multiplying by $\Theta^{-1}$ yields eq.(\ref{diffphi}).
\square

We are now ready to find the Schroedinger equation satisfied by  $\psi$. 
\begin{proposition}
The function $\psi(x, \Lambda,\mu)$ defined by eq.(\ref{baker1}) satisfies the linear second order differential 
equation
\begin{equation}
-\psi''(x, \Lambda,\mu) - [p_0^2 +  2(\langle E \vert \Phi \rangle)' ]\; \psi(x, \Lambda,\mu) = \Lambda^2  \psi(x, \Lambda,\mu)
\label{schroepsi}
\end{equation}
\end{proposition}
\proof
We have
$$
\psi(x, \Lambda,\mu) = U(x, \Lambda,\mu) - \langle {V}(x, \Lambda,\mu) \vert \Theta^{-1} \vert W(x) \rangle
= U(x, \Lambda,\mu) - \langle {V}(x, \Lambda,\mu)  \vert \Phi(x) \rangle
$$
Using the formula for $\langle {V}(x, \Lambda,\mu)  \vert$, we get
$$
\psi(x, \Lambda,\mu) = U \left(1 - \langle E' \vert {1\over Z^2  - K^2} \vert \Phi
 \rangle  \right)
+ U' \langle E \vert {1\over Z^2  - K^2} \vert \Phi \rangle
$$
Next, remembering that $U''(x, \Lambda,\mu) = - Z^2  U(x, \Lambda,\mu)$,  $\vert E''(x) \rangle = -K^2  \vert E(x) \rangle$,
we obtain
$$
\psi'(x, \Lambda,\mu) = -U\left(   \langle E \vert \Phi \rangle
+  \langle E' \vert {1\over Z^2  - K^2} \vert \Phi'
 \rangle \right) + U' \left( 1
+  \langle E \vert {1\over Z^2 - K^2} \vert \Phi'
 \rangle \right)
$$
and
\begin{eqnarray*}
\psi''(x, \Lambda,\mu) &=& -U\left(  Z^2  + \langle E \vert \Phi' \rangle+ (\langle E \vert \Phi \rangle)'
+  \langle E' \vert {1\over Z^2  - K^2} \vert \Phi''
 \rangle \right) \\
 && + U' \left( -  \langle E \vert \Phi \rangle
+  \langle E \vert {1\over Z^2  - K^2} \vert \Phi''
 \rangle \right)
\end{eqnarray*}
Using now the equation for $\vert \Phi'' \rangle$, we get eq.(\ref{schroepsi}).
\square

The potential  $T(x) = 2(\langle E \vert \Phi \rangle)'$ can also be written directly in terms of $\Theta$. In fact, we have
$$
\partial_x^2 \log \det \Theta = \partial_x\; {\rm Tr}\; \Theta^{-1} \Theta' = \partial_x \;\langle E \vert \Theta^{-1}  W \rangle =   
\partial_x \langle E \vert \Phi \rangle 
$$
hence
$$
T(x) = 2\partial_x \langle E \vert \Phi \rangle  = 2 \partial_x^2 \log \det \Theta(x) 
$$
The Schroedinger equation therefore also reads
\begin{equation}
\psi''(x, \Lambda,\mu) + [p_0^2 +  2\; \partial_x^2 \log \det \Theta  ]\; \psi(x, \Lambda,\mu) = -\Lambda^2  \psi(x, \Lambda,\mu)
\label{schroe1}
\end{equation}

In this formula both the potential and the function $\psi(x, \Lambda,\mu)$ are known. The potential 
therefore belongs to the class of exactly solvable potentials. It is strongly reminiscent 
of the formula for finite zones potentials \cite{No74, DuNo74, ItMa75, MKvM75}.
It can probably also be obtained by an infinite sequence of Darboux transformations
\cite{MaSa90}.

\bigskip

The parameter $\mu$ which enters the function $U(x,\Lambda,\mu)$ and the vector $\langle V(x,\Lambda,\mu ) \vert$
was, up to now, a free parameter. Eq.(\ref{baker1}) therefore provides two linearly independent solutions of eq.(\ref{schroepsi}). We now introduce the spectral curve by imposing the quasiperiodicity of $\psi(x,\Lambda,\mu )$.

\section{Bloch Waves and Spectral Curve.}

So far, $\psi(x, \Lambda,\mu)$ was defined on the interval $[ 0, 1 ]$. We extends its definition by imposing 
$$
\psi (x+1, \Lambda,\mu) = \mu \psi (x, \Lambda,\mu)
$$
This extension is continuous as we now show.
\begin{proposition}
$$
\psi(1, \Lambda,\mu) - \mu \psi(0, \Lambda,\mu) = 0
$$
\end{proposition}
\proof
This follows immediately from
$$
 W_k(1) = \mu_k^{-1} W_k(0),\quad
\Theta_{km}(1)=  \mu_k^{-1}\Theta_{km}(0) (-1)^{m}
\quad
 U(1, \Lambda,\mu) = \mu U(0, \Lambda,\mu)
 $$
\square

It is worth computing explicitly $\psi(0, \Lambda,\mu)$. In terms of the matrix $M$ introduced in eq.(\ref{defM}), we have
$$
\Theta_{km}(0) = W_k(0) M_{km} E'_{m}(0), \quad U(0, \Lambda,\mu) = {\sin Z \over Z}, \quad V_m(0, \Lambda,\mu) = U(0, \Lambda,\mu)  {1\over Z^2 - m^2 \pi^2}E'_m(0)
$$
It follows that
\begin{equation}
\psi(0, \Lambda,\mu) = {\sin Z \over Z} \left( 1 - \sum _m {\eta_m \over Z^2 - m^2\pi^2} \right)
= {\sin Z \over Z} \eta(Z)
\label{psi0}
\end{equation}
where $\eta_m$ and $\eta(Z)$ are defined in eqs.(\ref{defetam}, \ref{etaZ}). 
Notice that when $Z^2 = Z_k^2$, we have $\psi(0, \Lambda,\mu) = 0$, by definition of $\eta(Z)$.

We now turn to the derivative of $\psi(x, \Lambda,\mu)$.

\begin{lemma}
One has
$$
\psi'(1, \Lambda,\mu) - \mu \psi'(0, \Lambda,\mu) = \mu    \; \widetilde{\Gamma}(\Lambda ,\mu)
$$
\begin{equation}
\widetilde{\Gamma}(\Lambda ,\mu) = 
\mu + \mu^{-1} - 2\cos Z
- {\sin Z \over Z} \sum_{m=1}^\infty 
   {E'_m(0) \Phi'_m(0) - E'_m(1) \Phi'_m(1)\over Z^2 - m^2\pi^2 }
  \label{Gammatilde}
\end{equation}
\end{lemma}
\proof
We have
$$
U(1, \Lambda,\mu) = \mu {\sin Z \over Z},\quad U(0, \Lambda,\mu) =  {\sin Z \over Z}
$$
and
$$
U'(1, \Lambda,\mu) =\mu \cos Z -1, \quad U'(0, \Lambda,\mu) = \mu - \cos Z
$$
Using $E_k(1) = E_k(0)=0$, we get
$$
\psi'(1, \Lambda,\mu)=- U(1, \Lambda,\mu) \langle E'(1)\vert {1\over Z^2- K^2 } \vert \Phi'(1) \rangle 
+ U'(1, \Lambda,\mu)
$$
$$
\psi'(0, \Lambda,\mu)=- U(0, \Lambda,\mu) \langle E'(0)\vert {1\over Z^2- K^2 } \vert \Phi'(0) \rangle 
+ U'(0, \Lambda,\mu)
$$
From this the result follows.
\square

At this point it is tempting to identify the spectral curve as $\widetilde{\Gamma}(\Lambda, \mu) =0$.
However, this cannot be correct because the point $\Lambda = 0, \mu = \mu_0$ 
does not belong to it. 

\bigskip

We have to change the Schroedinger equation. The only possible modification  is at the 
edges. We consider therefore the equation
\begin{equation}
\psi''(x, \Lambda,\mu) + \Big[ p_0^2 + 2 \langle E \Phi \rangle' + H_0 \delta(x) \Big] \psi(x, \Lambda,\mu) = - \Lambda^2 \psi(x, \Lambda,\mu)
\label{schroe2}
\end{equation}
The bulk formula for $\psi(x, \Lambda,\mu)$ does not change. The continuity of $\psi(x, \Lambda,\mu)$ at $x=1$ still holds, but the derivative now has a discontinuity
$$
\int_{1-}^{1+}dx \;  \psi'' + H_0 \psi(1) = 0
$$
Using
$$
\psi(1, \Lambda,\mu) = \mu \;  \eta(Z) {\sin Z \over  Z}
$$
the Bloch condition becomes
$$
\psi'(1, \Lambda,\mu) - \mu \psi'(0, \Lambda,\mu) - H_0 \mu \; \eta(Z) = 0
$$
that is
$$
\mu + \mu^{-1} = 2 \cos Z + {\sin Z \over Z} \left(\sum_m {\Gamma_m \over Z^2 - m^2\pi^2}
- H_0 \eta(Z) \right)
$$
where we have set
\begin{equation}
\Gamma_m = E'_m(0) \Phi'_m(0) - E'_m(1) \Phi'_m(1)
\label{Gammambis}
\end{equation}

We now determine the coefficient $H_0$ by requiring that the curve passes through the  points
$p_0, \mu_0^{\pm 1}$. We  find
$$
H_0 = {1\over \eta(p_0)} \sum_m {\Gamma_m \over p_0^2 - m^2\pi^2}
$$
Hence the curve takes the form
$$
\mu + \mu^{-1} = 2 \cos Z + {\sin Z \over Z} \left(\sum_m {\Gamma_m \over Z^2 - m^2\pi^2}
- { \eta(Z)\over \eta(Z_0)}  {\Gamma_m \over p_0^2 - m^2\pi^2} \right)
$$

In order to compare with eq.(\ref{specurveGammaEta}), we must compute $\Gamma_m$.
We have
$$
W_k(0) = \mu_k {\sin Z_k \over Z_k}, \quad W_k(1) = {\sin Z_k \over Z_k}, \quad
W'_k(0) = 1-\mu_k \cos Z_k, \quad W'_k(1) = \cos Z_k - \mu_k
$$
and
$$
\Theta_{km}(0) = {W_k(0) E'_m(0) \over Z_k^2 - \pi^2 m^2}= W_k(0)M_{km} E'_m(0),\quad
\Theta_{km}(1) = {W_k(1) E'_m(1) \over Z_k^2 - \pi^2 m^2}= W_k(1)M_{km} E'_m(1)
$$
where $M$ is the matrix introduced in eq.(\ref{defM}). Since $\vert \Phi'(0) \rangle = \Theta^{-1}(0) \vert W'(0) \rangle$ and 
$\vert \Phi'(1) \rangle = \Theta^{-1}(1) \vert W'(1) \rangle$,  one has
$$
E'_m(0) \Phi'_m(0) = \sum_k M^{-1}_{mk} \; (\mu_k^{-1}
- \cos Z_k ) {Z_k \over \sin Z_k}, \quad
E'_m(1) \Phi'_m(1) = \sum_k M^{-1}_{mk} \;( \cos Z_k  - \mu_k) {Z_k \over \sin Z_k}
$$
hence
$$
\Gamma_m = E'_m(0) \Phi'_m(0)- E'_m(1) \Phi'_m(1) = \sum_k M^{-1}_{mk} \; {Z_k \over \sin Z_k} \gamma_k
$$
where we recall that $\gamma_k = \mu_k + \mu_k^{-1} - 2 \cos Z_k $. 
This is  exactly eq.(\ref{defGammam}) and shows that   we have recovered precisely the spectral  curve eq.(\ref{specurveGammaEta}). 

Finally, let us close this section by proving the

\begin{proposition}
The function $\psi(x,\Lambda,\mu)$ has no pole at the point $Z = Z_k, \mu_k^{-1}$
\end{proposition}
\proof
Here we restaure the factor $ {1 \over \prod_k  (1- Z_k^2/Z^2 )}$.
$$
U(x, \Lambda,\mu)\vert_{Z=Z_k , \mu_k^{-1}} = \mu_k^{-1} W_k(x), \Longrightarrow
V_{k'}(x, \Lambda,\mu)\vert_{Z=Z_k, \mu_k^{-1}} =\mu_k^{-1} \Theta_{kk'}(x)
$$
$$
\hskip -.5cm
\psi(x, \Lambda,\mu)\vert_{Z=Z_k, \mu_k^{-1}} \simeq  {1 \over \prod_k  (1- Z_k^2/Z^2 )}\;\;
\mu_k^{-1} \Big[ W_k(x) -  \Theta_{kk'}(x) 
\Theta_{k'k''}^{-1}(x)   W_{k''} (x)  \Big] = {\rm regular}
$$
\square

This shows that the same property for the eigenvector on the lattice has been preserved 
when taking the continuum limit.

\section{Perturbation theory.}

In the previous sections, we have manipulated determinants of infinite matrices quite freely. 
It is necessary now to investigate the conditions for the existence of the determinant $\det \Theta(x)$.
We recall  the free configuration eq.(\ref{freelambda}). In the scaled variables it reads
$$
Z_k^{(0)} = k\pi, \quad \mu_k^{(0)} = (-1)^k
$$
By construction, when $(Z_k, \mu_k) = (Z_k^{(0)}, \mu_k^{(0)})$, we have $\Theta(x) = {\rm Id}$, so that
$\det \Theta(x) = 1$. Clearly for $\det \Theta(x)$ to exist, we have to assume $(Z_k, \mu_k) \to (Z_k^{(0)}, \mu_k^{(0)})$ when $k \to  \infty$. Hence we set
\begin{equation}
Z_k = k\pi + \delta Z_k,\quad \mu_k = (-1)^k (1 + \delta \mu_k)
\label{smalldeviation}
\end{equation}
It is not difficult to see that to leading order in $(\delta Z_k, \delta \mu_k)$, we have
$$
W_k (x) = {1\over k\pi} ( \delta Z_k \cos k\pi x - \delta \mu_k \sin k\pi x )
$$
and this implies
\begin{eqnarray}
\Theta_{k,m}(x) &=& {2m \pi \over k(Z_k^2-\pi^2m^2)} \Big\{ \delta Z_k (m \cos k\pi x \cos m\pi x + k \sin k\pi x \sin m\pi x )\label{thetaasymp} \\
&&\hskip 3 cm + \delta \mu_k ( -m\sin k\pi x \cos m\pi x + k \cos k\pi x \sin m \pi x ) \Big\} \nonumber
\end{eqnarray}
Notice that when $m=k$ this formula gives that to leading order $\Theta_{k,k}(x) = 1$, as it should be.

A first consequence of these formulae is that if $\delta Z_k = 0, \delta \mu_k = 0$, beyond a certain 
index $k= k_{max}$, then for $k> k_{max}$ we have $W_k(x) = 0$, $\Theta_{k,k}(x) = 1$, and $\Theta_{k,m}(x) =0, \forall m\neq k$. As a result  only the first block of size $k_{max}\times k_{max}$ of the matrix $\Theta (x)$ plays a role and all the constructions of the previous sections reduce to finite size matrices and vectors. 

If however we want to retain  an infinite number  of modes in order to keep the field theoretical character 
of the model, one has to say something about the rate at which $\delta Z_k$ and $\delta \mu_k$ tend to zero when $k\to \infty$. Disregarding a finite number of possibly large $\delta Z_k, \delta \mu_k$ which
play no role in these convergence questions, we may assume that $\Theta (x)$ is given by eq.(\ref{thetaasymp}). As we have seen, it is of the form
$$
\Theta (x) = {\rm Id} + \widetilde{\Theta}(x)
$$
where $ \widetilde{\Theta}(x)$ is small.  In fact, bounding the trigonometric functions by $1$, we have
$$
\vert  \widetilde{\Theta}_{k,m} (x) \vert \le c {m\over k |k-m|} ( | \delta Z_k | + | \delta \mu_k | ), \quad m \neq k
$$
Since $|k-m| \geq 1$ when $k\neq m$, we may write as well
$$
\vert  \widetilde{\Theta}_{k,m} (x) \vert \le c {m\over k } ( | \delta Z_k | + | \delta \mu_k | ), \quad m \neq k
$$
It is not difficult to see that this formula is valid  also for $m=k$ (we have to adapt the constant $c$). It follows that
\begin{eqnarray}
\log \det \Theta (x) &=& {\rm Tr} \log ( 1 +  \widetilde{\Theta}(x) ) \nonumber \\
& \leq & \sum_{n=1}^\infty  {1\over n} {\rm Tr} | \widetilde{\Theta}(x)|^n
\leq   \sum_{n=1}^\infty  {c^n\over n}\left( \sum_k \Big( |\delta Z_k | + |\delta \mu_k | \Big) \right)^n
\label{boundingsum}
\end{eqnarray}
Hence a sufficient  condition for the existence of the determinant is that the series
$ \sum_k \Big( |\delta Z_k | + |\delta \mu_k | \Big)$ converges. This is achieved if 
\begin{equation}
 |\delta Z_k | + |\delta \mu_k |  < {c' \over k^{1+\epsilon}}, \quad \epsilon >0
 \label{existencecondition}
 \end{equation}
One can then adjust the constant $c'$ such that the series in eq.(\ref{boundingsum}) converges. The condition
eq.(\ref{existencecondition}) ensures that $\log \det \Theta (x)$ exists.  To build the potential $u(x)$
however, this  function has to be twice differentiable in $x$ and this may require stronger conditions on  $\epsilon$.

Now that we have found an expression for the potential $u(x)$ in terms of a countable set of variables
$Z_k, \mu_k$, we would like to check the Virasoro Poisson bracket directly. Recall that
$$
u(x) = p_0^2 + T(x) + H_0 \delta (x), \quad T(x) = 2 \partial_x \langle E \vert \Phi \rangle = \sum_n T_n e^{2i\pi x}
$$
Notice first that $T(x)$ has no Fourier component $T_0$:
$$
T_0 = \int_0^1 dx T(x) = 2 \int_0^1 dx  ( \langle E \vert \Phi \rangle )' = 2 ( \; \langle E(1) \vert \Phi(1) \rangle- \langle E(0) \vert \Phi(0) \rangle  \; ) =0
$$
where we used that $E_m(0)=E_m(1)=0$. The Fourier expansion of the potential $u(x)$ reads 
$$
u(x) = \sum_n L_n e^{2i\pi n x } = p_0^2 + \sum_n (T_n + H_0) e^{2i\pi n x}
$$
we must therefore  identify
\begin{equation}
L_0 =  p_0^2 + H_0  = p_0^2 + {1\over \eta(Z_0)} \sum_m {\Gamma_m \over p_0^2 - m^2\pi^2}
\label{defL0}
\end{equation}
and
$$
 L_n = T_n + H_0, \quad \Longrightarrow T_n = L_n - L_0 + p_0^2, \quad n\neq 0
$$
If $u(x)$ has Poisson bracket eq.(\ref{virpois}), the algebra of the $L_n$ reads
$$
\{ L_n, L_m\} = 8i\pi (n-m) L_{n+m} - 16i\pi^3 n^3 \delta_{n+m,0}
$$
The Poisson algebra for the $T_n$ is then closed:
$$
\{ T_n, T_m\} = 8i\pi (n-m)T_{n+m} - 8i\pi n T_n +8i\pi m T_m - 16i(\pi^3 n^3 - \pi p_0^2 n) \delta_{n+m,0}
$$
or, in a form that will be  useful later
\begin{equation}
\{ T(x), T(y)\} = 2 \delta'''(x-y) +4 (2 p_0^2 + T(x) +T(y)) \delta'(x-y) - 4T'(x) \delta(y) +4 T'(y) \delta(x)
\label{virasoro}
\end{equation}

In this section,  we consider the situation where all the variables $(Z_k, \mu_k)$
are close to the free configuration as in eq.(\ref{smalldeviation}), 
and we perform a perturbation theory in $\delta Z_k, \delta \mu_k$.
We have seen that to lowest order $\Theta^{(0)}(x) = {\rm Id}$ by construction. So, we can write the expansion
\begin{eqnarray*}
\Theta(x) &=& {\rm Id} + \Theta^{(1)}(x) + \Theta^{(2)}(x) + \cdots \\
 \vert W(x) \rangle &=&  \vert W^{(1)}(x) \rangle + \vert W^{(2)}(x) \rangle + \cdots
\end{eqnarray*}
where we have taken into account that $ \vert W^{(0)}(x) \rangle =0$. It follows 
that
$$
\vert \Phi (x) \rangle = \vert \Phi^{(1)}(x) \rangle + \vert \Phi^{(2)}(x) \rangle + \cdots
$$
where
$$
\vert \Phi^{(1)} \rangle  = \vert W^{(1)} \rangle, \quad
\vert \Phi^{(2)} \rangle = \vert W^{(2)} \rangle - \Theta^{(1)} \vert W^{(1)} \rangle,
\cdots
$$
To lowest order, we find easily
$$
T^{(1)}(x) = 2\langle E \vert \Phi^{(1)} \rangle' =4 \sum_k  k\pi (\delta Z_k   \cos 2k\pi x 
- \delta \mu_k  \sin 2k\pi x  )
$$
This shows in particular that $\delta Z_k$  and $\delta \mu_k$ are just the Fourier components of the potential in this first approximation. We see here clearly that for $T^{(1)}(x)$ to exist as a function (and not just as a distribution), we need $\epsilon > 1$ in eq.(\ref{existencecondition}).
The Poisson bracket eq.(\ref{poisson})  becomes to leading order
$$
\{ \delta Z_k, \delta \mu_k \} = 2k\pi  \left( 1 - {p_0^2 \over k^2\pi^2} \right) 
$$
To define modes independent of the zero mode $p_0$ we introduce 
$$
a_k = \alpha_k (\delta \mu_k - i \delta Z_k ), \quad
a_k^\dag = \bar{\alpha}_k (\delta \mu_k + i \delta Z_k )
$$
where the coefficients  $\alpha_k$, $\bar{\alpha}_k$ satisfy\footnote{It is known  \cite{GeNe} that the poles 
at $p_0^2 = k^2\pi^2$ are classical remnants of the zeroes of the Kac determinant.}
$$
\alpha_k \bar{\alpha}_k = {1\over 4\pi} { k^2 \pi^2 \over p_0^2 - k^2 \pi^2}
$$
With this choice, one has
$$
\{   a_k  , a^\dag_k\} = i k
$$
we can then rewrite
$$
T^{(1)}(x) = 2i \sum_k  k\pi \left( {a_k \over \alpha_k} e^{2ik\pi x} - {a_k^\dag \over \bar{\alpha}_k } e^{-2ik\pi x } \right)
$$
It is now straightforward to compute the Poisson bracket
\begin{eqnarray*}
\{ T^{(1)}(x), T^{(1)}(y) \}
&=&  2\delta'''(x-y) + 8 p_0^2 \delta'(x-y)
\end{eqnarray*}
This is the correct result for the Virasoro Poisson bracket  in  this approximation.
Notice that the term  $\delta'''(x-y)$  is exact already at this level. Higher order terms cannot contribute to it.

\bigskip

Next, we look at the conserved quantities. The leading terms in the expansions of
$\eta_m$ and $H_m$ are easy to find:  
$$
\eta_m \simeq 2m\pi\; \delta Z_m, \quad \Gamma_m \simeq 2 m^2\pi^2 (\delta \mu_m^2 +  \delta Z_m^2)
$$
To see it,  consider the defining relations of $\eta_m$
$$
\sum_m {\eta_m \over Z_k^2 - \pi^2 m^2} = 1, \forall k
$$
When $Z_k$ is given by eq.(\ref{smalldeviation}), the dominant term in the above sum  
is $m=k$. The equation becomes $\eta_k / (2 k\pi \delta Z_k) = 1$. 
The same argument starting with the equation of the spectral curve, eq.(\ref{specurveH0Hm}), 
taken at the point  $Z_k, \mu_k$ which belongs to it, yields the formula for $H_m$. 
Remark that the condition eq.(\ref{existencecondition}) ensures that the sums in the definition of the function $\eta(Z)$, eq.(\ref{etaZ}), or in the definition of the spectral curve, eq.(\ref{specurveH0Hm}),  are convergent.

Written with the oscillators $a_m, a_m^\dag$, we find
$$
H_m = 8\pi (p_0^2 - \pi^2 m^2) a_m^\dag a_m, \quad
H_0 = 8\pi \sum_m a_m^\dag a_m
$$
It is clear that the $H_m$ are in involution
 at this order. As we see, in first approximation,
  the dynamical system reduces to a set of decoupled 
 harmonic oscillators. The generator $L_0$ is given by
 $$
 L_0 = p_0^2 + 8\pi \sum_m a_m^\dag a_m
 $$
 It is easy to  verify that
$$
\{ L_0, T^{(1)}(x) \}_0 = - 4 \partial_x T^{(1)}(x)
$$

These perturbative arguments are  good indications that $u(x)$ indeed satisfies the Virasoro
Poisson bracket. Clearly we will not go very far in perturbation and we now look for a more formal 
proof of this fact. For that purpose, we need some preparation.

\section{Some identities.}
Before computing Poisson brackets to check the Virasoro algebra, we collect a number 
of useful identities.  We start with a formula for the   inverse matrix $\Theta^{-1}(x)$. It has the same form as $\Theta(x)$.
\begin{proposition}
Let us define
\begin{equation}
\langle F \vert = \langle E \vert \Theta^{-1}
\label{defF}
\end{equation}
Then, we can write
\begin{equation}
\Theta^{-1}_{mk} = {
\Phi_m F'_{k} - \Phi'_m  F_{k} 
\over
 Z_{k}^2  - \pi^2 m^2 }
 \label{thetainverse}
\end{equation}
\end{proposition}
\proof
 Multiplying 
eq.(\ref{eqint}) on both sides by $\Theta^{-1}$, we get
$$
\Theta^{-1} {\cal Z}^2 - K^2 \Theta^{-1} = \Theta^{-1} \vert W \rangle \langle E' \vert \Theta^{-1}
-  \Theta^{-1} \vert W' \rangle \langle E \vert \Theta^{-1}
$$
and so
$$
\Theta^{-1}_{mk} = {
 ( \Theta^{-1} \vert W' \rangle)_m  (\langle E \vert \Theta^{-1})_{k} -
 (\Theta^{-1} \vert W \rangle)_m (\langle E' \vert \Theta^{-1})_{k}
 \over
 \pi^2 m^2 - Z_{k}^2  }
$$
But
$$
\Theta^{-1} \vert W' \rangle = \vert \Phi' \rangle + \langle E \vert \Phi \rangle 
\vert \Phi \rangle, 
\quad 
 \langle E' \vert \Theta^{-1} = \langle F' \vert +  \langle E \vert \Phi \rangle 
\langle F \vert
$$
Plugging into the above formula, we obtain eq.(\ref{thetainverse}).
\square

\begin{proposition}
The vector $\langle F \vert$ satisfies a set of differential equations. 
\begin{equation}
\langle F'' \vert + 2 (\langle E \vert \Phi \rangle)' \langle F \vert + \langle F \vert {\cal Z}^2 = 0
\label{diffF}
\end{equation}
\end{proposition}
\proof
The proof is the same as for $ \vert \Phi \rangle$.

From this we easily deduce
$$
(\Theta^{-1})' = - \vert \Phi \rangle \langle F \vert
$$
which can also be proved using the similar property of $\Theta$.
Let us define
$$
A_k(x) = \sum_{m} {E_{m} (x) \Phi_{m}(x) \over Z_k^2 - m^2\pi^2},
\quad
B_k(x) = \sum_{m} {E'_{m} (x) \Phi_{m}(x) \over Z_k^2 - m^2\pi^2}
$$
$$
C_k(x) = \sum_{m} {E_{m} (x) \Phi_{m}(x) \over (Z_k^2 - m^2\pi^2)^2},
\quad 
D_k(x) = \sum_{m} {E'_{m} (x) \Phi_{m}(x) \over (Z_k^2 - m^2\pi^2)^2}
$$

\begin{proposition}
We have the identity
\begin{equation}
(1-B_k)W_k + A_k W'_k  = 0
\label{id4}
\end{equation}
\end{proposition}
\proof
This is just a rewriting of $\vert W \rangle = \Theta \vert \Phi \rangle  $ using eq.(\ref{thetax})
for $\Theta(x)$.
\square

\begin{proposition}
The following two identities hold
\begin{eqnarray}
(1-B_k+A'_k)F_k - A_k F'_k &=& 0
\label{id1} \\
(B'_k+ Z_k^2 A_k)F_k +(1- B_k) F'_k &=& 0
\label{id2}
\end{eqnarray}
\end{proposition}
\proof
The first identity is a rewriting of $\langle F \vert = \langle E \vert \Theta^{-1}$ using eq.(\ref{thetainverse})
for $\Theta^{-1}(x)$. The second identity is just a  rewriting of
$\langle F' \vert =  \langle E' \vert \Theta^{-1}  - \langle E \vert \Phi \rangle \langle F \vert $.
\square
The above two identities form a linear system for $F_k$ and $F'_k$.  Its compatibility 
 implies the following:
\begin{proposition}
\begin{equation}
(1-B_k)^2 + A_k B'_k - A'_k B_k + A'_k + Z_k^2 A_k^2 = 0
\label{id3}
\end{equation}
\end{proposition}
Let us define 
\begin{equation}
\widetilde{F}_k = W'_k (1- B_k) + W''_k A_k
\label{defFtilde}
\end{equation}
An important consequence of eqs.(\ref{id1},\ref{id2}) is
\begin{proposition}
The functions $F_k(x)$ and $\widetilde{F}_k(x)$ are proportional.
\begin{equation}
F_k(x)  = - {\zeta_k \over \mu_k \gamma_k}  \widetilde{F}_k (x)
\label{FFtilde}
\end{equation}
The proportionality  coefficient is written in this specific way  
for later convenience. The quantity $\gamma_k$ is the one defined in eq.(\ref{gammak}).
\end{proposition}
\proof
Let us compute the Wronskian
\begin{eqnarray*}
Wr(F_k, \widetilde{F}_k) &=& F_k\Big(W_k''(1-B_k)- W'_kB'_k + W'''_k A_k + W''_k A'_k \Big) 
 - F'_k \Big(W'_k(1-B_k) + W''_k A_k \Big) \\
&=& - \Lambda^2_k W_k \Big( (1-B_k + A'_k) F_k - A_k F'_k \Big) 
 - W'_k \Big( (B'_k + Z_k^2 A_k) F_k + (1-B_k)F'_k \Big) 
= 0
\end{eqnarray*}
\square

\begin{proposition}
The following quantities are constants independent of $x$
\begin{equation}
\eta_m =  \Phi_m E'_m - \Phi'_m E_m - < E \Phi> \Phi_m E_m  
- \sum_{p\neq m} {1\over \pi^2(p^2-m^2)}(\Phi_m \Phi'_p -\Phi'_m \Phi_p )
(E'_m E_p - E_m E'_p ) 
\label{etamwronskian}
\end{equation}
The quantities $\eta_m$ defined in eq.(\ref{etamwronskian}) are in fact  the same as the ones introduced in eq.(\ref{defetam}).
\end{proposition}
\proof
To prove the first statement, just take the derivative with respect to $x$ and use eq.(\ref{diffphi}).
To prove the second statement, use the $\eta_m$ defined in eq.(\ref{etamwronskian}) to rewrite 
 eq.(\ref{id3}) as
\begin{equation}
\sum_m {\eta_m  \over (Z^2_k -m^2\pi^2)}=1
, \quad \forall \; k
 \label{etambis}
\end{equation}
which is exactly the same as eq.(\ref{defetam}).
\square
A straightforward consequence   is the "trace" formula that will be  useful later:
\begin{equation}
\langle E \Phi' \rangle - \langle E' \Phi \rangle + \langle E \Phi \rangle^2 =
-\sum_m \eta_m
\label{sumetam}
\end{equation}

We now compute the coefficients $\zeta_k$ appearing in eq.(\ref{FFtilde})
\begin{proposition}
The coefficients $\zeta_k$ in eq.(\ref{FFtilde}) are determined by the set of equations
\begin{equation}
\sum_{k} {\zeta_k \over Z^2_{k} - m^2\pi^2 } = 1, \quad 
\forall \;  m
\label{id8}
\end{equation}
These equations are dual to eq.(\ref{etambis}).
\end{proposition}
\proof
Start with
$$
\langle F \vert W \rangle = \langle E \vert \Theta^{-1} \vert W \rangle = \langle E \vert \Phi \rangle
$$
Using eq.(\ref{id4}, \ref{defFtilde},\ref{FFtilde} ), we have
\begin{equation}
F_k W_k = -{\zeta_k \over \mu_k \gamma_k}(- W_k'^2 + W_k''W_k) A_k = \zeta_k A_k
\label{FW0}
\end{equation}
hence
$$
\langle F \vert W \rangle = \sum_k  \zeta_k  A_k = \sum_m \left(\sum_{k} {\zeta_k \over Z^2_k -m^2\pi^2}\right)
E_{m} \Phi_{m} =  \langle E \vert \Phi \rangle = \sum_{m} E_{m} \Phi_{m}
$$
Since this has to hold for all $x$,   the only possibility is eq.(\ref{id8}).
\square

We collect below a few more identities of the type of eq.(\ref{FW0}) that will be important later
\begin{proposition}
\begin{eqnarray}
F_k W_k &=& \zeta_k \; A_k \label{FW} \\
F_k W'_k &=& - \zeta_k \;  (1-B_k) \label{FdW} \\
F'_k W_k &=&  \zeta_k \; (1-B_k + A'_k) \label{dFW} \\
F'_k W'_k &=&  \zeta_k \;  (B'_k +Z_k^2  A_k) \label{dFdW} 
\end{eqnarray}
\end{proposition}

Next, we relate the $\eta_m$ and $\zeta_k$
\begin{proposition}
The following relation holds:
\begin{eqnarray*}
\sum_{m} {\eta_m   \over (Z^2_k -m^2\pi^2)^2}
 ={1\over \zeta_k}
\end{eqnarray*}
\end{proposition}
\proof
We start from
$$
\sum_m  \Theta_{km} \Theta^{-1}_{mk'} = \delta_{kk'}
$$
When $k=k'$ this gives
$$
W_k F'_k D_k - W_k F_k (D'_k - A_k + Z_k^2 C_k ) - 
W'_k F'_k C_k + W'_k F_k (C'_k - D_k ) = 1
$$
or using eqs.(\ref{FW}-\ref{dFdW})
\begin{eqnarray*}
\zeta_k \Big\{ 2 D_k (1-B_k) + D_k A'_k - D'_k A_k + A_k^2 -2 Z_k^2 C_k A_k
- C_k B'_k - C'_k (1-B_k) \Big\} = 1
\end{eqnarray*}
Expanding this formula using $(p \neq m)$
\begin{eqnarray*}
 {1\over (Z_k^2 - \pi^2 m^2)^2(Z_k^2 - \pi^2 p^2)}&=&
-{1\over \pi^2(p^2-m^2)}{1\over (Z_k^2 - \pi^2 m^2)^2} \\
&&- {1\over \pi^4(p^2-m^2)^2}\left\{ {1\over (Z_k^2 - \pi^2 m^2)}
- {1\over (Z_k^2 - \pi^2 p^2)}\right\}
\end{eqnarray*}
we get the result with $\eta_m$ represented  by eq.(\ref{etamwronskian}).
\square

An immediate and important consequence is an expansion of the function 
$\eta(Z)$ near $Z^2 = Z_k^2$ 
 \begin{equation}
\eta(Z) =   (Z^2 - Z_k^2){1\over 
 \zeta_k } + \cdots
 \label{etaprime}
  \end{equation}

Returning to the formula for $\psi(x, \Lambda, \mu)$, we can write it as
$$
\psi(x, \Lambda, \mu) = {\mu - e^{-i Z} \over 2iZ } w(x,Z)
-  {\mu - e^{iZ} \over 2i Z } w^*(x,Z)
$$
where we define
\begin{eqnarray}
w(x,Z) &=& e^{i Z x}
 \left(1 - \langle E' \vert {1\over Z^2  - K^2} \vert \Phi
 \rangle
+  \langle E \vert {iZ \over Z^2  - K^2} \vert \Phi \rangle   \right) 
\label{w}\\
w^* (x,Z) &=& e^{-i Z x}
 \left(1 - \langle E' \vert {1\over Z^2  - K^2} \vert \Phi
 \rangle 
-  \langle E \vert {iZ \over Z^2  - K^2} \vert \Phi \rangle  \right)
\label{w*}
\end{eqnarray}
The functions $w(x,Z)$ and $w^*(x,Z)$ are defined on the Riemann sphere with a puncture at $\infty$.
One can compute the Wronskian
$$
w' w^* - {w^*}' w = 2iZ  \eta(Z)
$$
This Wronskian vanishes precisely when $Z^2 = Z^2_k$.
Hence, when  $Z = Z_k$,  $w(x,Z_k)$ becomes proportional
to $w^*(x,Z_k)$. Indeed we have
\begin{eqnarray*}
w(x,Z_k) &=& e^{iZ_k x} (1-B_k + iZ_k A_k) \\
w^*(x,Z_k) &=& e^{-iZ_k x} (1-B_k - iZ_k A_k)
\end{eqnarray*}
Then  eq.(\ref{id4})  can be rewritten as
\begin{equation}
w(x,Z_k) = {\alpha_k^{(+)} \over \alpha_k^{(-)} }
w^*(x,Z_k)
\label{ww*}
\end{equation}
where
\begin{equation}
\alpha_k^{(\pm)} = 1-\mu_k e^{\pm i Z_k}, \quad 
 \alpha_k^{(+)} \alpha_k^{(-)} = \mu_k \gamma_k
 \label{alpha+alpha-}
\end{equation}

The function $\widetilde{F}_k(x)$ introduced in eq.(\ref{defFtilde})
is solution of eq.(\ref{diffF}). It is not difficult to see that the other
solution of this equation is
$$
\widetilde{G}_k = A_k W'_k + 2Z_k (W_k D_k - W'_k C_k) + x \widetilde{F}_k
$$

In terms of $w(x,Z),w^*(x,Z)$ defined in eqs.(\ref{w},\ref{w*}), we have
\begin{eqnarray}
\widetilde{F}_k &=& {1\over 2} (\alpha_k^{(-)} w\vert_{Z_k} + \alpha_k^{(+)} w^*\vert_{Z_k})
= \alpha_k^{(-)} w\vert_{Z_k} = \alpha_k^{(+)} w^*\vert_{Z_k}
\label{Fw}\\
\widetilde{G}_k &=& -{i\over 2} ( \alpha_k^{(-)} \partial_Z w\vert_{Z_k} - \alpha_k^{(+)} \partial_Z w^*\vert_{Z_k} )
\label{Gw}
\end{eqnarray}
This will play an important role below.

\section{Virasoro algebra.}
We are now ready to compute the Poisson bracket $\{ T(x), T(y) \}$. The result is
precisely  the algebra of the $T_n= L_n - L_0$, eq.(\ref{virasoro}).

\begin{proposition}
Let $T(x)=2\partial_x \langle E(x) \vert \Phi(x) \rangle$, and $X(x), Y(x)$ be two arbitrary test functions. Then we have
\begin{eqnarray}
\left\{ \int_0^1 dx X(x) T(x) ,  \int_0^1 dy Y(y) T(y) \right\} &=&- \int_0^1 dx ( X''' Y - X Y''') 
 - 4  \int_0^1 dx (X' Y - X  Y')(p_0^2 + T) \nonumber \\
&& \hskip -4cm 
+ 4 \int_0^1 dx X(x) \delta(x) \int_0^1 dy Y(y) T'(y) 
 - 4 \int_0^1 dx X(x) T'(x) \int_0^1 dy Y(y) \delta(y) 
 \label{virproof}
\end{eqnarray}
\end{proposition}

The proof is rather long and we will split it into several lemmas.
Since $\vert \Phi \rangle = \Theta^{-1} \vert W \rangle$,
we have
\begin{eqnarray*}
\{ \vert \Phi \rangle_1,  \vert \Phi \rangle_2 \} &=&
\Theta_1^{-1}\Theta_2^{-1}\Big[
 \{ \Theta_1, \Theta_2 \} \vert \Phi_1 \rangle \vert \Phi_2 \rangle   
  -  \{ \Theta_1, W_2 \} \vert \Phi_1 \rangle   
   -   \{ W_1, \Theta_2 \} \vert \Phi_2 \rangle    +  \{ W_1, W_2 \} \Big]
\end{eqnarray*}
where the index $1,2$ refers to the customary tensor notation.
In this expression, all Poisson brackets can be computed explicitly. 
Using the fact that rows with different indices in $\Theta$ and $W$ Poisson commute and using only eq.(\ref{id4}),
we arrive at
\begin{equation}
\{  \Phi_m(x) ,  \Phi_n(y)  \} = - 2 \sum_k
\Theta_{mk}^{-1}(x)\Theta_{nk}^{-1}(y)
 {\sin Z_k \over Z_k \gamma_k} \left(1- {p_0^2\over Z_k^2}\right)
\Big[ \widetilde{F}_k(x) \widetilde{G}_k(y) - \widetilde{F}_k(y) \widetilde{G}_k(x) \Big]
\label{phiphi}
\end{equation}
where $\gamma_k $ are defined in eq.(\ref{gammak}).
Multypliying by $E_m(x) E_n(y)$ and remembering that $\langle E \vert \Theta^{-1} = \langle F \vert $
we get
\begin{equation}
\{ \langle E(x) \vert \Phi(x) \rangle ,  \langle E(y)\vert  \Phi(y) \rangle \} =
2\sum_k {\zeta_k^2 \sin Z_k \over Z_k \gamma_k^3 \mu_k^2 }
 \left(1- {p_0^2\over Z_k^2}\right) 
\times \Big( {\cal A}_k(y) {\cal B}_k(x) - {\cal A}_k(x) {\cal B}_k(y) \Big)
\label{PoiTT}
\end{equation}
where
$$
{\cal A}_k(x) =  \widetilde{F}_k^2(x), \quad 
{\cal B}_k(x) = \widetilde{F}_k(x) \widetilde{G}_k(x)
$$

Using eqs.(\ref{w},\ref{w*}), we can write
\begin{eqnarray*}
{\cal A}_k (x) &=& \mu_k \gamma_k \; w(x,Z) w^*(x,Z) \vert_{Z = Z_k} \\
{\cal B}_k(x) &=& {\mu_k \gamma_k \over 2i} ( w^*(x,Z) \partial_Z w(x,Z) - w(x,Z) \partial_Z w^*(x,Z) )
\vert_{Z = Z_k}
\end{eqnarray*} 

The strategy to evaluate the right hand side of eq.(\ref{PoiTT}) is to rewrite it as a sum over the residues of certain poles of a function on the Riemann $Z$-sphere. This sum can then be transformed as a sum over the residues of the other poles (there will be none in our case) plus a integral over a small circle at infinity surrounding an essential singularity. 
This last integral can then be evaluated using  the known asymptotics of the function.
Let us define
\begin{equation}
a(Z) = {Z \over 2i \sin Z } e^{-iZ}, \quad
b(Z) = - {Z \over 2i \sin Z } \cos Z, \quad
c(Z) = {Z \over 2i \sin Z } e^{iZ}
\label{abc}
\end{equation}
and introduce the functions
\begin{eqnarray*}
\Omega_1 &=& w(x,Z) w^*(y,Z) - w^*(x,Z) w(y,Z) \\
\Omega_2 &=& a(Z) w(x,Z) w(y,Z)  \\
&& +b(Z) (w(x,Z) w^*(y,Z) + w^*(x,Z) w(y,Z) )\\
&&+ c(Z) w^*(x,Z) w^*(y,Z)
\end{eqnarray*}
These functions are defined on the Riemann $Z$-sphere  have poles at the points $\pm m\pi$
and have an essential singularity at infinity. Let
$$
\Omega = \Omega_1 \Omega_2
$$
and  recall the definition of $\eta(Z)$ eq.(\ref{etaZ}).

\begin{lemma}
\begin{equation}
\sum_{\pm Z_k}{\rm Res} {\Omega \over\eta^2(Z)}\left(1- {p_0^2\over Z^2}\right)
 = - 2\sum_k  {\zeta_k^2 \sin Z_k\over Z_k
\gamma_k^3 \mu_k^2} \left(1- {p_0^2\over Z_k^2}\right) \times 
\Big({\cal A}_k(y) {\cal B}_k(x) - {\cal A}_k(x) {\cal B}_k(y)\Big)
\label{theequation}
\end{equation}
\end{lemma}
\proof
The factor $\eta^2(Z)$ introduces double poles at $\pm Z_k$ because $\eta(Z_k) = 0$. However
using eq.(\ref{ww*}), we see immediately that  $\Omega_1 \vert_{\pm Z_k} = 0$, 
so that the poles are in fact simple. Remembering eq.(\ref{etaprime}), we have
\begin{eqnarray*}
\sum_{\pm Z_k}{\rm Res} {\Omega \over\eta^2(Z)} \left(1- {p_0^2\over Z^2}\right)  &=& \sum_k {\zeta_k^2 \over 4Z_k^2} \left(1- {p_0^2\over Z_k^2}\right)  \Big( \partial_Z \Omega \vert_{Z_k} + \partial_Z \Omega \vert_{-Z_k} \Big) 
\end{eqnarray*}
We need to compute 
$\partial_Z \Omega \vert_{\pm Z_k} = \partial_Z \Omega_1 \vert_{\pm Z_k} \Omega_2  \vert_{\pm Z_k}$.
Evaluating at $Z_k$ gives
\begin{eqnarray*}
\partial_Z \Omega\vert_{Z_k} &=&
{2i\over \mu_k^2 \gamma_k^2} \left(a(Z_k) {\alpha_k^+\over \alpha_k^-} + 2 b(Z_k) + c(Z_k) {\alpha_k^-
\over \alpha_k^+} \right) \Big({\cal A}_k(y) {\cal B}_k(x) - {\cal A}_k(x) {\cal B}_k(y)\Big)
\end{eqnarray*}
while using $w(-Z_k) = w^*(Z_k)$, $\partial_Z w \vert_{-Z_k}
= - \partial_Z w^* \vert_{Z_k}$, we also have
\begin{eqnarray*}
\partial_Z \Omega\vert_{-Z_k} &=&
{2i \over \mu_k^2\gamma_k^2}\left(c(-Z_k) {\alpha_k^+\over \alpha_k^-} + 2 b(-Z_k) + a(-Z_k) {\alpha_k^-
\over \alpha_k^+} \right) \Big({\cal A}_k(y) {\cal B}_k(x) - {\cal A}_k(x) {\cal B}_k(y)\Big)
\end{eqnarray*}
The result follows from the identities
\begin{eqnarray*}
a(Z_k) {\alpha_k^+\over \alpha_k^-} + 2b(Z_k) +c(Z_k){\alpha_k^-
\over \alpha_k^+}  &=&  2i{Z_k \sin Z_k \over \gamma_k} \\
c(-Z_k){\alpha_k^+\over \alpha_k^-} + 2 b(-Z_k) + a(-Z_k) {\alpha_k^-
\over \alpha_k^+}  &=&  2i{Z_k \sin Z_k \over \gamma_k} 
\end{eqnarray*}
\square

Next we have to examine the poles at $\pm m\pi$ in the expression
$$
{\Omega \over\eta^2(Z)}\left(1- {p_0^2\over Z^2}\right)
$$
We rewrite $\Omega_1$ and $\Omega_2$ as
\begin{eqnarray*}
\Omega_1 &=& (w(x) - w^*(x) )w^*(y) - (w(y) - w^*(y) )w^*(x) \\
 \Omega_2 &=& {Z \over 4i\sin Z} \times 
\Big\{  (w^*(x) - w(x) ) (e^{iZ} w^*(y) - e^{-iZ} w(y) )
+  \\
&&\hskip 3cm
+ (w^*(y) - w(y) ) (e^{iZ} w^*(x) - e^{-iZ} w(x) ) \Big\}
\end{eqnarray*}
Recalling the formula eq.(\ref{w}, \ref{w*}) for $w(x,Z)$ and $w^*(x,Z)$, we see
that when $Z = 0$, we have $w = w^*$ so that  $\Omega_1 = 0(Z)$ and
$\Omega_2 = 0(Z^2)$. Hence we have no pole at $Z = 0$.
When $Z = \pm \pi m + \epsilon$
$$
w = {1\over \epsilon} w_{\pm m}^{(-1)} + w_{\pm m}^{(0)} + \cdots, \quad
w^* = {1\over \epsilon} w_{\pm m}^{*(-1)} + w_{\pm m}^{*(0)} + \cdots
$$
with
$$
 w_{\pm m}^{(-1)}(x) =  w_{\pm m}^{*(-1)}(x) =  \mp \pi m \Phi_m(x) \\
$$
Because the two leading terms are the  same,
both $w^*(x,Z) - w(x,Z)$ and  $e^{iZ} w^*(x,Z) - e^{-iZ} w(x,Z)$ are regular. So  $\Omega_1$
and $\Omega_2$ both behaves like $1/\epsilon$. 
Since $1/\eta^2(Z)$ behaves like $\epsilon^2$, the whole thing is in fact regular.

We come to the  conclusion that  everything happens at infinity. We want to compute
\begin{equation}
\left\{ \int dx X(x) T(x),  \int dy Y(y)  T(y) \right\} = 4\int_0^1 dx X(x) \int_0^1 dy Y(y) \int_{C_\infty} dZ
\left(1- {p_0^2\over Z^2}\right){1\over\eta^2(Z)} \partial_x \partial_y \Omega 
\label{tocompute}
\end{equation}
Let
\begin{eqnarray*}
\widetilde{w}(x,Z) &=& \eta^{-1/2}(Z) \; w(x,Z) \\
\widetilde{w}^*(x,Z) &=&  \eta^{-1/2}(Z) \;  w^*(x,Z)
\end{eqnarray*}
The wronskian of $\widetilde{w}(x,Z)$ and $\widetilde{w}^*(x,Z)$ is $2iZ$ and therefore 
these functions coincide with the Baker--Akhiezer functions which are usually introduced
in the pseudo-differential appproach to the KdV hierarchy (see e.g. \cite{BaBeTa03}).
At $Z \simeq \infty$, we have
\begin{eqnarray*}
\widetilde{w}(x,Z)&=&
e^{iZ x}\left( 1 - {\omega(x) \over i Z } + {\omega'(x) + \omega^2(x) 
 \over 2( i Z)^2 }+ \cdots  \right) \\
\widetilde{w}^*(x,Z)&=&
e^{-iZ x}\left( 1 + {\omega(x) \over i Z } + {\omega'(x) + \omega^2(x) 
 \over 2( i Z)^2 }+ \cdots  \right)
\end{eqnarray*}
where we have set
$$
\omega(x) =  \langle E \Phi \rangle (x)
$$
Eq.(\ref{sumetam}) is needed to verify this formula.
Using theses  asymptotic forms, we find
\begin{eqnarray*}
\partial_x \left( \widetilde{w}^2(x,Z) \right) &=& \left( \sum_{n=-\infty}^1 A_n(x) (iZ)^n \right) e^{2iZ x} \\
\partial_x \left( \widetilde{w}^{*2}(x,Z) \right)  &=& \left( \sum_{n=-\infty}^1 (-1)^n A_n(x) (iZ)^n \right) e^{-2iZ x} \\
\partial_x \left( \widetilde{w}(x,Z) \widetilde{w}^*(x,Z) \right) &=&
\sum_{n=-\infty}^{-1} C_{2n}(x) (iZ)^{2n}
\end{eqnarray*}
where
$$
A_1 = 2,\quad A_0 = -4 \omega(x),\quad A_{-1} = 4 \omega^2(x), \quad
A_{-2} = -{8\over 3} \omega^3 -2 \int^x \omega'^2,
\quad A_{-3} = {4\over 3} \omega^4 + 4 \omega \int^x \omega'^2
$$
$$
C_{-2}(x) = \omega''(x)
$$

Consider the term proportional  to $b(Z)$ in  eq.(\ref{tocompute}):
\begin{eqnarray*}
4\int_{C_\infty} {dZ \over 2i\pi} b(Z)\left( 1 - {p_0^2\over Z^2}\right) \int dx dy 
(X(x) Y(y) - X(y) Y(x))
 \partial_x \widetilde{w}^2(x,Z) \partial_y  \widetilde{w}^{*2}(y,Z) 
\end{eqnarray*}

\begin{lemma}
Let us define
$$
I_p(z) =  \int_{C_\infty} {dZ \over 2i\pi} b(Z) ( i Z)^p
 e^{2iZ z} 
$$
We have
$$
I_{-1}(z) = -{1\over 2}\delta(z), \quad 
I_{0}(z) = -{1\over 4}\delta'(z), \quad 
 I_{1}(z) = -{1\over 8}\delta''(z), \quad 
 I_{2}(z) = -{1\over 16}\delta'''(z)
 $$
$$
I_{-n}(z)= -{2^{n-3}\over (n-2)!} \epsilon(z) z^{n-2}, \quad n \geq 2
$$
\end{lemma}
\proof
It is clear that
$$
\partial_z I_p(z) = 2 I_{p+1}(z)
$$
hence we can determine all the $I_p(z)$ recursively.  For $p \geq 0$ this is done 
by successively differentiating $I_{-1}(z)$ which is easy to calculate
 $$
 I_{-1}(z) = \int_{C_\infty} {dZ \over 2i\pi} b(Z){1\over i Z}
 e^{2iZ z}  = {1\over 2} \int_{C_\infty} {dZ \over 2i\pi} {\cos Z \over \sin Z} 
 e^{2iZ z} = -{1\over 2} \sum_{n\in Z} e^{2in\pi z} = -{1\over 2} \delta(z)
$$
For $p \leq -2$ we have to successively integrate $I_{-1}(z)$. For this we need boundary conditions which are provided by
\begin{equation}
\int_{C_\infty} {dZ \over 2i\pi} b(Z) Z^{-p}  = 0,
\quad p \geq 2
\label{Ibp}
\end{equation}
This is because
\begin{eqnarray*}
\int_{C_\infty} {dZ \over 2i\pi} b(Z)\sum_{p=2}^\infty \zeta^p Z^{-p}  &=& 
-{\zeta \over 2 i} \int_{C_\infty} {dZ \over 2i\pi} {\cos Z \over \sin Z}
\sum_{p=2}^\infty \zeta^{p-1} Z^{-p + 1}  \\
&&\hskip -3cm
=-{\zeta^2\over 2 i} \int_{C_\infty} {dZ \over 2i\pi} {\cos Z \over \sin Z}
{1 \over Z - \zeta} = {\zeta^2\over 2 i} \left( \cot \zeta - {1\over \zeta}
- \sum_{n=0}^\infty {2\zeta \over \zeta^2 - n^2\pi^2 } \right) = 0 \end{eqnarray*}
\square

Denoting
$$
F_p(x,y) = \sum_{m+n=p} (-1)^m A_n( x)A_m(y )
$$
we get
 \begin{eqnarray*}
4\int_0^1 dx \int_{0}^1dy  \left(X(x )Y(y)
-X(y )Y(x) \right)\sum^{p=2}_{-\infty} F_p(x,y)
(I_p(x-y) + p_0^2 I_{p-2}(x-y) )
 \end{eqnarray*}
In  this expression, we separate  the terms with a $\delta(x-y)$ function or its derivative
which will lead to local terms ($L_b$) and  the non local terms ($NL_b$) with are proportional to $\epsilon(x-y)$.
\begin{eqnarray}
L_b &=& 4 \int dx dy (X(x) Y(y) - X(y) Y(x)) \Big( F_2 I_2  + F_1  I_1
+( F_0+ p_0^2 F_2)  I_0 +( F_{-1} + p_0^2 F_1) I_{-1}  \Big) 
\nonumber\\
NL_b &=&4 \int dx dy (X(x) Y(y) - X(y) Y(x)) 
 \sum_{n=0}^\infty (F_{-2-n} + p_0^2 F_{-n}) I_{-n-2}
 \label{nlb}
\end{eqnarray}
  We have
$$
F_2(x,y) =  -4, \quad F_1(x,y) = 8 \left( \omega (x) -   \omega(y) \right)  
$$
$$
F_0(x,y) = -8 \left(  \omega(x) -   \omega(y) \right)^2, \quad
F_{-1}(x,y) = {16\over 3}(\omega(x)-\omega(y))^3 
+ 4 \int_y^x \omega'^2
$$
The local terms are
\begin{eqnarray*}
L_b &= &4\int_0^1 dx \int_0^1 dy (X(x)Y(y)-X(y)Y(x)) \left\{ {1\over 4 } \delta'''(x-y) + p_0^2 \delta'(x-y)+ \right.\\
&& \left. -(\omega(x)-\omega(y)) \delta''(x-y) 
- 2 (\omega(x)-\omega(y))^2 \delta'(x-y) -{1\over 2} F_{-1}(x,y) \delta(x-y)
\right\}
\end{eqnarray*}
The last two terms obviously vanish and what remains is
\begin{eqnarray*}
L_b= \int_0^1 dx \left\{ - (X''' Y - X Y''' ) - 4( (X' Y - X Y' )(p_0^2 + 2 \omega' )\right\}
\end{eqnarray*}

\begin{lemma}
The non local term eq.(\ref{nlb}) is identically zero.
\end{lemma} 
\proof
The non local term reads
\begin{eqnarray*}
NL_b &=& -2\int_0^1 dx \int_0^1 dy (X(x) Y(y)-X(y)Y(x)) \epsilon(x-y) \\
&& \hskip 3cm 
\left[ \sum_{n=0}^\infty 
\Big(F_{-2-n}(x,y) + p_0^2 F_{-n}(x,y) \Big){2^{n}\over (n)! } (x-y)^{n} \right]
\end{eqnarray*}
The first sum  is just the coefficient of $(iZ)^{-2}$ in the formal expansion of $\partial \widetilde{w}^2(x,Z) \partial \widetilde{w}^{*2}(y,Z)$ while the second sum is the coefficient of $(iZ)^{0}$.
Hence we have
$$
NL_b = 2\int_0^1 dx \int_0^1 dy (X(x) Y(y)-X(y)Y(x)) \epsilon(x-y) \int_{C_\infty} {dZ \over 2i\pi} (Z - p_0^2 Z^{-1}) \; \partial \widetilde{w}^2(x,Z) \partial \widetilde{w}^{*2}(y,Z)
$$
The above expression is zero in the following sense. Let us write
$$
 \int_{C_\infty} {dZ \over 2i\pi} (Z - p_0^2 Z^{-1}) \; \partial \widetilde{w}^2(x,Z) \partial \widetilde{w}^{*2}(y,Z)
 = \sum_{i=1}^\infty { (y-x)^i\over i !} 
 \int_{C_\infty} {dZ \over 2i\pi} (Z - p_0^2 Z^{-1}) \; \partial \widetilde{w}^2(x,Z) \partial^{i+1} \widetilde{w}^{*2}(x,Z)
 $$
We will show that all the integrals around $C_\infty$ in the right hand side are identically zero.

Since the function  $ \widetilde{w}(x,Z)$ satisfies the Schroedinger equation, 
its square $\widetilde{w}^2(x,Z)$ satisfies a third order differential equation
$$
{\cal D} \widetilde{w}^2(x,Z) = - 4 Z^2 \partial \widetilde{w}^2(x,Z), \quad {\cal D} = \partial^3 + 8 \omega' \partial + 4 \omega''
$$
Let us introduce a pseudo differential operator $\Phi$ such that
$$
\widetilde{w}^2(x,Z) = \Phi e^{2iZ x}
$$
Then
$$
{\cal D} = \partial \Phi \partial^2 \Phi^{-1}
$$
Since ${\cal D}$ is anti self-adjoint,  we also have
$$
{\cal D} = -{\cal D}^* = \Phi^{*-1} \partial^2  \Phi^* \partial
$$
it follows that $\widetilde{w}^{*2}(x,Z)$ which is solution of
$$
(-{\cal D}^*) \widetilde{w}^{*2}(x,Z) = -4 Z^2 \partial  \widetilde{w}^{*2}(x,Z) 
$$
can be written as
$$
 \widetilde{w}^{*2}(x,Z) = \partial^{-1} \Phi^{*-1} \partial e^{-2iZ x}
 $$
 Hence
 $$
 \partial \widetilde{w}^2 = \partial \Phi e^{2iZ x} , \quad  \partial^{i+1} \widetilde{w}^{*2} = \partial^i \Phi^{*-1} \partial e^{-2iZ x} 
$$
Finally, we have to compute
\begin{eqnarray*}
 \int_{C_\infty} {dZ \over 2i\pi} (Z - p_0^2 Z^{-1})\; ( \partial \widetilde{w}^2(x,Z)  )  \partial^{i+1} \widetilde{w}^{*2}(x,Z) &=& 
  \int_{C_\infty} {dZ \over 2i\pi} (Z - p_0^2 Z^{-1})\;  ( \partial  \Phi e^{2iZ x}
  )  \partial^i \Phi^{*-1} \partial e^{-2iZ x} \\
  &=& {-1\over 2i} \int_{C_\infty} {dZ \over 2i\pi}  \; ( \partial  \Phi e^{2iZ x}
  ) \partial^i   \Phi^{*-1}( \partial^2 + 4 p_0^2)e^{-2iZ x}
\end{eqnarray*}
 We recall the formula (see e.g.\cite{BaBeTa03})
 $$
 \int_{C_\infty} {dZ \over 2i\pi} (D e^{iZ x} ) (F e^{-iZ x}) = {\rm Res}_\partial (DF^*)
 $$
 where ${\rm Res}_\partial$ is Adler's residue \cite{Adler79}. So our expression is equal to
 $$
 {\rm Res}_\partial (  \partial  \Phi  (\partial^2 + 4p_0^2)\Phi^{-1} \partial^i) =  {\rm Res}_\partial  ( ({\cal D} +4 p_0^2\partial)\partial^i ) = 0
 $$
because $ ({\cal D} + 4p_0^2\partial)\partial^i $ is a differential operator.
\square

Consider next the term proportional to $a(Z)$  in eq.(\ref{tocompute}).
\begin{eqnarray*}
4\int_0^1 dx \int_0^1 dy \Big(X(x) Y(y) - X(y) Y(x)\Big) \int_{C_\infty} {dZ \over 2i\pi} a(Z) \left( 1 - {p_0^2\over Z^2}\right)\partial_x  \widetilde{w}^2(x,Z)
 \partial_y \Big( \widetilde{w}(y,Z) \; \widetilde{w}^*(y,Z) \Big) 
\end{eqnarray*}
\begin{lemma}
Let us define
$$
J_p(x) = \int_{C_\infty} {dZ \over 2i\pi} a(Z) (iZ)^p e^{2iZ x},
\quad p = -1,-2, \cdots
$$
One has
$$
J_{-1}(x) = {1\over 2} \delta(x), \quad J_{-p}(x) = 2^{p-2} {x^{p-2}\over (p-2)!} ( \epsilon(x) - 1), \quad 
p \geq 2
$$
\end{lemma}
\proof
One has
$$
\partial_x J_p(x) = 2 J_{p+1}(x)
$$
The calculation of $J_{-1}(x)$ is easy. Next, we need boundary conditions to determine
the other $J_p$ by integration
\begin{eqnarray*}
\sum_{n=2}^\infty (i\zeta)^n J_{-n}(0) &=& {\zeta^2 \over 2i}  \int_{C_\infty} {dZ \over 2i\pi} {e^{-iZ} \over \sin Z} 
{1\over Z -  \zeta } \\
&=&  {\zeta^2 \over 2i} \left[ -{e^{-i\zeta } \over \sin \zeta} + \sum_{n\in Z} {1\over \zeta - n\pi} \right] = 
 {\zeta^2 \over 2i} \left[ -{e^{-i\zeta } \over \sin \zeta}  + \cot \zeta \right]  = {\zeta^2 \over 2}
\end{eqnarray*}
\square
It follows that all the non local terms containing $J_p(x)$ for $p \leq -2$ vanish 
 when $0 < x < 1$. The $a(Z)$ term is
$$
L_a = 4\int_0^1 dx \int_0^1 dy \Big(X(x) Y(y) - X(y) Y(x)\Big)  A_1(x) C_{-2}(y)J_{-1}(x)
$$
or
$$
L_a = 4 \int_0^1 dx X(x) \delta(x)  \int_0^1 dy Y(y) \omega''(y) - 4\int_0^1 dx Y(x) \delta(x)  \int_0^1 dy X(y) \omega''(y)
$$
Finally, the term in $c(Z)$ is just equal to the $a(Z)$ one and double it.

Putting everything together, we arrive at eq.(\ref{virproof}). In the course of this proof, we have 
shown the  identities
$$
 \int_{C_\infty} {dZ \over 2i\pi} (Z - p_0^2 Z^{-1})\; ( \partial \widetilde{w}^2(x,Z)  )  \partial^{i+1} \widetilde{w}^{*2}(x,Z) = 0, \quad \forall i \geq 0
 $$
These are quartic identities on the coefficients of $\widetilde{w}(x,Z)$. Of course, we also have the 
quadratic relations of Hirota and Sato that were interpreted by Sato as Pl\"{u}cker relations defining the infinite Grassmannian,  allowing to give a precise definition of $\tau$-functions. The above relations are quartic relations on $\tau$-functions analogous to the quartic relations on Riemann Theta functions.

\section{Poisson bracket  $\{ L_0, u(y)\}$.}

In the previous section, we have  obtained the Poisson bracket for the 
generators $T_n = L_n-L_0$. We now have to reintroduce $L_0$ and check that it has the correct
Poisson brackets. The candidate for $L_0$ was given in eq.(\ref{defL0}). Let us recall it:
$$
L_0 = p_0^2 + {1\over \eta(p_0)} \sum_k {\Gamma_k \over p_0^2 - k^2\pi^2}
$$
where
$$
\Gamma_k = E'_k(0)\Phi'_k(0) - E'_k(1) \Phi'_k(1) ) 
$$
and
$$
 \eta (Z) = 1 - \sum_m {\eta_m \over Z^2 - m^2\pi^2 } 
= 1 - \sum_m { E'_m(0)\Phi_m(0) \over Z^2 - m^2\pi^2 }
$$
where we have used eq.(\ref{etamwronskian}), evaluated at $x=0$, to express 
$\eta_m$. 
\begin{proposition}
We have the following Poisson bracket:
\begin{equation}
\left\{ L_0, \int_0^1 dy Y(y) u(y) \right\} = -4 \int_0^1dy Y(y) \Big[(L_0 - p_0^2) \delta'(y) + T'(y) \Big]
\label{L0proof}
\end{equation}
This shows that $\{L_0, \cdot \} $ acts on $u(y)$ as $\partial_y$,  as it should be.
\end{proposition}

Again, the proof is long and we will split it  into several lemmas.
We need to compute $\{ \Phi_m(x), T(y) \}$.  and $\{ \Phi_m'(x), T(y) \}$
for $x = 0$ and $x=1$.

Multiplying eq.(\ref{phiphi})  by $E_n(y)$ and remembering that 
$\langle E \vert \Theta^{-1} = \langle F \vert $ and 
$F_k(x) = -{\zeta_k \over \gamma_k \mu_k} \widetilde{F}_k(x) $, we get
\begin{eqnarray}
\{  \Phi_m(x) ,  \langle E(y)\Phi(y)\rangle  \} &=&
\label{phimux} \\
&&\hskip -4cm
 =-2 \Phi_m(x)\sum_k  {\sin Z_k \over Z_k}
 {\zeta_k^2( 1 - p_0^2 Z_k^{-2} )\over \mu_k^2 \gamma_k^3 (Z_k^2 - 
\pi^2 m^2)}   \Big[ \widetilde{F}'_k(x)\widetilde{F}_k(x)  \widetilde{F}_k(y)\widetilde{G}_k(y) - \widetilde{F}^2_k(y)  \widetilde{F}'_k(x)\widetilde{G}_k(x) \Big] \nonumber \\
&& \hskip -3.5cm
+ 2 \Phi'_m(x)\sum_k   {\sin Z_k \over Z_k}
 {\zeta_k^2( 1 - p_0^2 Z_k^{-2} )\over \mu_k^2 \gamma_k^3 (Z_k^2 - 
\pi^2 m^2)}   \Big[ \widetilde{F}^2_k(x)  \widetilde{F}_k(y)\widetilde{G}_k(y) - \widetilde{F}^2_k(y)  \widetilde{F}_k(x)\widetilde{G}_k(x) \Big] \nonumber
\end{eqnarray}

As before, this can be expressed in terms  of $\Omega_1$ and $\Omega_2$.  We find
\begin{eqnarray*}
\{  \Phi_m(x) ,  \langle E(y)\Phi(y)\rangle  \} &=& -
\sum_k {\rm Res}_{\pm Z_k} {( 1 - p_0^2 Z^{-2} ) \over \eta^2(Z)(Z^2 
- \pi^2 m^2)} \Omega_1 ( \Phi_m(x) \partial_x \Omega_2  -\Phi'_m(x) \Omega_2 )
\end{eqnarray*}
This formula is the starting point  to begin the computation of  $\{  \eta(p_0) ,  \langle E(y)\Phi(y)\rangle  \}$.
Setting $x=0$ in eq.(\ref{phimux}), the term proportional to  $\Phi'_m(x)$ vanishes
because  $F_k(0)=0$. We are left with ($x=0$, but we keep it for a while)
\begin{eqnarray}
\{  \Phi_m(x) ,  \langle E(y)\Phi(y)\rangle  \} &=&- \Phi_m(x)
\sum_k {\rm Res}_{\pm Z_k} {( 1 - p_0^2 Z^{-2} ) \over \eta^2(Z)(Z^2 
- \pi^2 m^2)} \Omega_1   \partial_x \Omega_2 
 \label{aux1}
\end{eqnarray}
Multiplying by ${ E'_m(0) \over  (p_0^2 -m^2\pi^2)}$ and summing over $m$, 
 in the right hand side appears the sum
\begin{eqnarray*}
-\sum_m { E'_m(0)\Phi_m(0) \over  (p_0^2 -m^2\pi^2) (Z_k^2 - \pi^2 m^2)}
&=&- \sum_m { \eta_m \over  (p_0^2 - \pi^2 m^2) (Z_k^2 - \pi^2 m^2)} \\
&=& {-1\over Z_k^2 - p_0^2 }\sum_m  \left({ \eta_m \over p_0^2 - m^2\pi^2 }
- { \eta_m \over  Z_k^2 - \pi^2 m^2} \right) \\
&=& {1\over Z_k^2 - p_0^2 }  \eta(p_0)
\end{eqnarray*}
where we used that $\sum_m  { \eta_m \over  Z_k^2 - \pi^2 m^2}  =  1, \quad \forall k$.
The factor $1/(Z_k^2 - p_0^2)$ cancels with the factor $( 1 - p_0^2 Z_k^{-2} )$  in eq.(\ref{aux1})
and we are left with
\begin{eqnarray*}
\{ \eta(p_0), \langle E \Phi \rangle (y) \} &=&-   \eta(p_0)
\sum_k {\rm Res}_{\pm Z_k} {1 \over \eta^2(Z)Z^2 } \Omega_1   \partial_x \Omega_2 
\end{eqnarray*}
Finally
\begin{equation}
\left\{ \eta(p_0), \int_0^1 dy Y(y) T(y) \right\} =  2\eta(p_0) 
\int_0^1 dy Y(y) 
\int_{C_\infty} {dZ \over 2i\pi} {1\over Z^2} \partial_y (\widetilde{\Omega}_1
\partial_x \widetilde{\Omega}_2 )
\label{etaT1}
\end{equation}
where we used the asymptotics
expressions for $\widetilde{w}$ and $\widetilde{w}^*$ inside $\Omega_1$ and $\Omega_2$, hence removing the $\eta^2(Z)$ factor. The last step consists in evaluating the integral around $C_\infty$.
The reult is:
\begin{lemma}
\begin{equation}
\left\{ \eta(p_0), \int_0^1 dy Y(y) T (y) \right\} = 4 \eta(p_0)
\int_0^1 dy Y(y) \delta'(y)
\label{etaT}
\end{equation}
\end{lemma}
\proof
Let us consider the integral over $C_\infty$ in eq.(\ref{etaT1}). 
The term containing $b(Z)$ can be written as
\begin{eqnarray*}
L_b &=&\int_{C_\infty} {dZ\over 2i\pi} {b(Z) \over Z^2 } \times \Big\{ (\widetilde{w}(x,Z) \widetilde{w}^{*\prime}(x,Z) - \widetilde{w}'(x,Z) \widetilde{w}^*(x,Z)  )\partial_y (\widetilde{w}(y,Z) \widetilde{w}^*(y,Z)) + \\
&& \hskip 3cm
 \widetilde{w}(x,Z) {\widetilde{w}}'(x,Z) \partial_y  \widetilde{w}^{*2}(y,Z)
-   \widetilde{w}^*(x,Z) \widetilde{w}^{*\prime}(x,Z) \partial_y  {\widetilde{w}}^2(y,Z) \Big\}
\end{eqnarray*}
On the first line, we recognize the wronskian of $\widetilde{w}(x,Z)$ and $\widetilde{w}^*(x,Z)$, which is just 
equal to $-2i Z$. Since  $\partial_y (\widetilde{w}(y,Z) \widetilde{w}^*(y,Z))=Z^{-2}S'_2(y) + \cdots $ this
 term  vanish by eq.(\ref{Ibp}).
Hence 
\begin{eqnarray}
L_b &=& {1\over 2}\int_{C_\infty} {dZ\over 2i\pi} {b(Z) \over Z^2 } \Big\{ 
\partial_x {\widetilde{w}^2}(x,Z) \partial_y  \widetilde{w}^{*2}(y,Z)
-  \partial_x \widetilde{w}^{*2}(x,Z) \partial_y  {\widetilde{w}}^2(y,Z)  \Big\} \\
&=& -\sum_{p=-2}^\infty F_{-p}(x,y) I_{-p-2}(x-y)  \nonumber\\
&=& \delta'(y-x) + {1\over 2} \epsilon(x-y) \int_{C_\infty} {dZ\over 2i\pi} {1\over Z }
\partial_x \widetilde{w}^2 (x,Z) \partial_y  \widetilde{w}^{*2}(y,Z)
\label{bterm}
\end{eqnarray}
The $\epsilon(x-y)$ term is zero because
\begin{eqnarray*}
\int_{C_\infty} {dZ\over 2i\pi} {1\over Z} \partial^{i+1} \widetilde{w}^2(x,Z) \partial \widetilde{w}^{*2}(x,Z) 
&=& \int_{C_\infty} {dZ\over 2i\pi} {1\over Z}( \partial^{i+1} \Phi e^{2iZ x})
(\Phi^{*-1} \partial e^{-2iZ x}) \\
&=&  \int_{C_\infty} {dZ\over 2i\pi} ( \partial^{i+1} \Phi e^{2iZ x})
(\Phi^{*-1} e^{-2iZ x}) \\
&=& {\rm Res}_\partial (\partial^{i+1} \Phi \cdot \Phi^{-1} ) = 0 
\end{eqnarray*}

\bigskip

Next, we consider the $a(Z)$ term. It reads
\begin{eqnarray*}
L_a &=& \partial_y \int_{C_\infty} {dZ\over 2i\pi} {a(Z) \over Z^2} (\widetilde{w}(x,Z) \widetilde{w}^*(y,Z)-\widetilde{w}^*(x,Z) \widetilde{w}(y,Z))
\widetilde{w}'(x,Z) \widetilde{w}(y,Z) \\
&=&   \int_{C_\infty} {dZ\over 2i\pi} {a(Z) \over Z^2}
\left({1\over 2} \partial_x \widetilde{w}^2(x,Z) \partial_y (\widetilde{w}(y,Z) \widetilde{w}^*(y,Z)) \right. \\
&& \hskip 2.5cm -{1\over  2}( \widetilde{w}^*(x,Z)  \widetilde{w}'(x,Z) +  \widetilde{w}^{*'}(x,Z) \widetilde{w}(x,Z)) \partial_y \widetilde{w}^2(y,Z) \\
&&
\left. \hskip 2.5cm -{1\over  2}( \widetilde{w}^*(x,Z)  \widetilde{w}'(x,Z) -  \widetilde{w}^{*'}(x,Z) \widetilde{w}(x,Z)) \partial_y
\widetilde{w}^2(y,Z) \right)
\end{eqnarray*}
Again, the last term is the wronskian and so
\begin{eqnarray*}
L_a&=&   \int_{C_\infty} {dZ\over 2i\pi} {a(Z) \over Z^2}
\left( -iZ  \partial_y( \widetilde{w}^2(y,Z) +
{1\over 2} \partial_x \widetilde{w}^2(x,Z)\partial_y (\widetilde{w}(y,Z) \widetilde{w}^*(y,Z)) \right. \\
&&\left. \hskip 3cm -{1\over  2} \partial_x( \widetilde{w}^*(x,Z)  \widetilde{w}(x,Z)) \partial_y( \widetilde{w}^2(y,Z))\right) 
\end{eqnarray*}
The first term is
\begin{eqnarray*}
 \int_{C_\infty} {dZ\over 2i\pi}  a(Z) (iZ)^{-p-1}A_{-p}(y) e^{2iZ y}
 &=& \sum_{p=-1}^\infty A_{-p}(y) J_{-p-1}(y) \\
&=&
  {1\over 2}\delta'(y) -\omega(y) \delta(y) + (\epsilon(y)-1) \sum_{p=1}^\infty A_{-p}(y)
 {(2y)^{p-1}\over (p-1)! }
 \end{eqnarray*}
 The last sum is zero  because $0 < y < 1$. The  $\delta(y)$ term vanishes because 
 $\omega(0)=0$. 
The second term is
\begin{equation}
{1\over 2}  \int_{C_\infty} {dZ\over 2i\pi}  {a(Z) \over Z^2} 
A_{-p}(x) C_{-q}(y) (iZ)^{-p-q} e^{2iZ x} = 
-{1\over 2} (\epsilon(x) -1) \Big( A_1(x) C_{-2}(y) (2x) + \cdots \Big)
\label{aterm}
\end{equation}
This vanishes when $x=0$.
The third term is
\begin{eqnarray*}
-{1\over 2}  \int_{C_\infty} {dZ\over 2i\pi}  {a(Z) \over Z^2} 
A_{-p}(y) C_{-q}(x) (iZ)^{-p-q} e^{2iZ y} = 
{1\over 2} (\epsilon(y) -1) \Big( A_1(y) C_{-2}(x) (2y) + \cdots \Big)
\end{eqnarray*}
and this vanishes when $0 < y < 1$.

Finally, it is easy to see that the $c(Z)$ term is equal to the $a(Z)$ one.
Putting everything together, we get eq.(\ref{etaT}).
\square

The last result we need is the following:
\begin{lemma}
\begin{equation}
\left\{ \sum_m { E_m'(0) \Phi'_m(0) - E_m'(1) \Phi'_m(1)\over p_0^2 - \pi^2 m^2} ,
\int_0^1 dy Y(y) T(y) \right\}= - 4 \eta(p_0)\int_0^1 dy Y(y) T'(y)
\label{GammaT}
\end{equation}
\end{lemma}
\proof
Taking the derivative with respect to  $x$ of eq.(\ref{phimux})
and remembering that $F_k(0)=F_k(1)=0$, the remaining terms are (there is a cancellation in the 
$\Phi'_m(x)$ term)
\begin{eqnarray}
\{  \Phi'_m(x) ,  \langle E(y)\Phi(y)\rangle  \} &=&- 2 \Phi_m(x)\sum_k  {\sin Z_k \over Z_k}
 {\zeta_k^2(1-p_0^2 Z_k^{-2})\over \mu_k^2 \gamma_k^3 (Z_k^2 - 
\pi^2 m^2)}\times
\nonumber \\
&&
 \partial_x \Big[ \widetilde{F}'_k(x)\widetilde{F}_k(x)  \widetilde{F}_k(y)\widetilde{G}_k(y) - \widetilde{F}^2_k(y)  \widetilde{F}'_k(x)\widetilde{G}_k(x) \Big] \nonumber
\end{eqnarray}
where it is understood that $x = 0$ or $x=1$. By exactly the same argument as before
\begin{eqnarray}
\{ \sum_m{E'_m(x) \Phi'_m(x)\over p_0^2 - \pi^2 m^2} , \int_0^1dy Y(y)  T(y)  \} &=&
 2 \eta(p_0) \int_0^1 dy Y(y) \partial_y \partial_x  \int_{C_\infty} {dZ \over 2i\pi} {1\over Z^2 \eta(Z)} \Omega_1
\partial_x \Omega_2 \nonumber
\end{eqnarray}
Hence, we just have to take the derivative with respect to $x$ of the previous result, before setting 
$x=0$ or $x=1$. At $x=0$ we  get
$$
2 \eta(p_0) \int_0^1 dy Y(y) \Big[ - \delta''(y) +4   \omega''(y) \Big]
$$
The $\delta''(y)$ term comes from eq.(\ref{bterm}) while the  second term comes from 
 eq.(\ref{aterm}) doubled by the $c(Z)$ contribution. At 
$x=1$ only the periodic $\delta''(y)$ remains. Taking the difference
we obtain eq.(\ref{GammaT}).
\square

Putting everything together, we arrive at eq.(\ref{L0proof}), and this finishes our proof that 
$u(x)$ does satisfy the Virasoro Poisson bracket.

\section{Conclusion.}

We have succeeded to take the continuum limit in the formulae expressing 
the dynamical variables of the Volterra model in terms of the separated variables.
This yields exactly solvable potentials and formulae for the Virasoro generators 
of a rather unusual type. Still, we were able to check that they have the correct 
Poisson brackets. Of course the most interesting thing now is to try to  quantize 
this approach. As a first step, a semiclassical analysis along the lines of 
\cite{Smi} should be very enlightening. The full quantum theory however
may reserve some surprise. The bracket eq.(\ref{poisson}) being in fact 
an ordinary quadratic bracket, it is natural to quantize it with Weyl type
commutation relations. This opens up the possibility  of phenomena as 
those advocated in \cite{Ba90, Fa}.

{\bf Acknowledgements.} This work was supported in part by the European Network
ENIGMA, Contract number : MRTN-CT-2004-5652.

 \end{document}